\documentclass[%
 reprint,
 amsmath,amssymb,
 aps,
 prresearch  % Add this to specify PRR
]{revtex4-2}

\usepackage{graphicx}% Include figure files
\usepackage{dcolumn}% Align table columns on decimal point
\usepackage{bm}% bold math
\usepackage{amsmath,amssymb,amsfonts}
\usepackage{hyperref}
\usepackage{xcolor}

\usepackage{xurl}  % Vor biblatex laden!
\usepackage{microtype}  % WICHTIG: Verbessert Umbrüche deutlich!
\usepackage{float}
\usepackage{dsfont}
\usepackage{braket}
\usepackage[normalem]{ulem}
\usepackage{comment}
\usepackage{xcolor}        % for \textcolor

\usepackage[colorinlistoftodos,textsize=footnotesize]{todonotes}

\begin{document}

\preprint{APS/123-QED}

\title{Quantum algorithm for one-quasiparticle excitations in the thermodynamic limit via cluster-additive block diagonalization
}% Force line breaks with \\

\author{Sumeet}
 \email{sumeet.sumeet@fau.de}

\author{M. H\"ormann}%
% \email{max.hoermann@fau.de}
\author{K. P. Schmidt}
% \email{kai.phillip.schmidt@fau.de}
\affiliation{Department of Physics, Friedrich-Alexander-Universit\"at Erlangen-N\"urnberg (FAU), 91058 Erlangen, Germany
}%

\date{\today}

\begin{abstract}
We propose a quantum algorithm for computing one-quasiparticle excitation energies in the thermodynamic limit by combining numerical linked-cluster expansions (NLCEs) and the variational quantum eigensolver (VQE). Our approach uses VQE to block-diagonalize the cluster Hamiltonian through a single-unitary transformation. This unitary is then postprocessed using the projective cluster-additive transformation (PCAT) to ensure cluster additivity, a key requirement for NLCE convergence. We benchmark our method on the transverse-field Ising model (TFIM) in one and two dimensions, and with longitudinal field, computing one-quasiparticle dispersions in the high-field polarized phase. We compare two cost function classes, trace minimization and variance based, demonstrating their effectiveness with the Hamiltonian variational ansatz (HVA). For pure TFIM, $\lceil N/2 \rceil$ layers of HVA suffice: NLCE+VQE matches exact diagonalization. For TFIM with longitudinal field, where parity symmetry breaks and PCAT becomes essential, both $\lceil N/2 \rceil$ and $N$ layers of HVA converge with increasing cluster size, with $N$ layers providing improved accuracy. Our results establish PCAT as a cluster-additive framework that extends variational quantum algorithms to excited-state calculations in the thermodynamic limit via NLCE. While demonstrated with VQE, the PCAT postprocessing approach, which requires only low-energy eigenspace information, applies to any quantum eigenstate preparation method.

\end{abstract}

\maketitle

\section{Introduction}
\label{sec:intro}

Understanding elementary excitations is central to the study of quantum many-body systems. Such excitations govern low-temperature thermodynamics and long-range dynamical behavior, and are directly accessible via spectroscopic techniques such as inelastic neutron or light scattering. Consequently, developing accurate and efficient methods to compute the excitation spectrum is a major objective in condensed matter physics, as well as in quantum chemistry and quantum simulation.

Recent years have witnessed significant progress in hybrid quantum-classical algorithms, particularly in using the variational quantum eigensolver (VQE) \cite{Peruzzo2014a} to compute ground-state properties of strongly correlated systems. These algorithms are particularly relevant for noisy intermediate-scale quantum devices \cite{Preskill2018}, where limited qubit coherence times and gate fidelities preclude error-corrected quantum computation. Originally designed for quantum chemistry applications \cite{Peruzzo2014a, OMalley2016}, VQE has been increasingly applied to condensed matter models, including studies of quantum magnets \cite{Kandala2017, Hempel2018}, spin liquids \cite{wiersema_exploring_2020}, frustrated systems \cite{Joris, Li2023}, quantum critical points \cite{Watanabe2024, Wang2023}, and our recent numerical linked-cluster expansion (NLCE)+VQE approach for ground-state energies \cite{Sumeet2024}. Recent no-go theorems have highlighted trainability challenges for variational quantum algorithms under certain assumptions \cite{McClean2018, Holmes2022}. However, systematic exploration with problem-adapted ansätze remains essential for developing successful strategies, and our work contributes to this by exploring VQE for degenerate subspaces within the NLCE framework using physically motivated initialization.

VQE has proven successful for ground-state calculations, providing a robust method to minimize the energy expectation value on finite quantum systems. Extending these approaches to compute excited states is an important next step, with multiple strategies having been proposed. These include variational quantum deflation \cite{Higgott2019}, quantum equation of motion \cite{Ollitrault2020}, subspace-search VQE (SSVQE) \cite{Nakanishi2019}, and other ensemble and penalty-based methods \cite{Benavides-Riveros2022, Guo2024}. Some of these techniques (such as SSVQE) can handle degenerate subspaces, though they are primarily designed for extracting individual eigenstates rather than constructing effective models for low-energy physics. 

In many quantum lattice models, elementary excitations can be interpreted as quasiparticles: stable or long-lived excitations carrying a well-defined momentum $\vec{k}$ and energy $\omega(\vec{k})$. Constructing an effective quasiparticle Hamiltonian provides a compact description of these excitations through a few local coupling constants, reveals the underlying interaction structure, and connects directly to experimental observables.

Classical approaches to construct such effective Hamiltonians rely on block-diagonalization methods in quantum many-body physics, such as the Schrieffer-Wolff transformation \cite{Schrieffer-Wolff}, continuous unitary transformations ~\cite{Wegner1994, glazek1993renormalization, knetter2000perturbation}, and the contractor renormalization group \cite{CORE}, which aim to isolate quasiparticle sectors by transforming the full Hamiltonian into a block-diagonal form where each block conserves the number of quasiparticles. Perturbative linked-cluster expansions \cite{Oitmaa2006, Gelfand1996} provide a systematic way to reach the thermodynamic limit by combining such block-diagonalization transformations with cluster expansion techniques. Perturbative approaches are limited to perturbative regimes. However, all of the previously mentioned methods can also be formulated non-perturbatively. Then, NLCEs \cite{irving1984linked, TANG2013557, Rigol2006, Coester2015a}, meaning replacing the perturbative with the exact result on a cluster, can be used to overcome the perturbative limitation. However, the exponential scaling still limits accessible cluster sizes.

While NLCE combined with exact diagonalization (ED) has been successful for both ground states and excitations, our recent work has demonstrated NLCE+VQE for ground-state energies \cite{Sumeet2024}, a method where VQE is used as a cluster solver instead of ED for each finite cluster in NLCE. However, a quantum or hybrid quantum-classical analog of these NLCE methods for constructing cluster-additive effective Hamiltonians of excited states does not exist yet. Recently, VQE-based approaches have been applied to extract quasiparticle excitations from finite periodic systems \cite{Balents}, but these methods target individual eigenstates on finite periodic clusters rather than constructing cluster-additive transformations suitable for thermodynamic-limit extrapolation via NLCE. Extending the NLCE framework to excitations with VQE as a cluster solver presents unique challenges due to higher dimensionality of excitation subspaces on finite clusters, which requires ensuring cluster additivity, a property not automatically satisfied by standard VQE block-diagonalization unitaries.

In this work, we propose a hybrid quantum-classical method to construct an effective one-quasiparticle (1QP) Hamiltonian in the thermodynamic limit. Our approach combines VQE with NLCEs and generalizes the NLCE+VQE method that we have recently introduced for ground-state energies \cite{Sumeet2024}. Extending this framework to excitations presents significant challenges: Unlike the ground-state energy (a scalar quantity), 1QP excitations form a multi-dimensional subspace on finite clusters, requiring careful treatment to maintain the cluster additivity essential for NLCE convergence.

To ensure the transformation satisfies cluster additivity, which is essential for NLCE convergence, we proceed in two stages. Within each cluster of the NLCE, we employ a VQE-based cost function to approximate a unitary transformation that decouples the ground and 1QP subspaces. From this, we construct the projective cluster-additive transformation (PCAT) \cite{hormann_projective_2023}, which guarantees cluster additivity, a property that the transformation factorizes for disconnected clusters \cite{knetter2003structure}.
We use a single unitary to decouple the full low-energy subspace (ground state and 1QP states) from higher excitations, rather than targeting individual eigenstates. As a result, a single VQE optimization simultaneously yields the entire $N$-dimensional 1QP block on a finite cluster, from which the full dispersion $\omega(\vec{k})$ in the thermodynamic limit is recovered after NLCE embedding, in contrast to approaches that obtain quasiparticle energies one momentum at a time. This block-diagonalization approach exploits the fact that energy gaps within the low-energy subspace are typically small, while gaps to higher excitations are large, facilitating efficient decoupling. We use the Hamiltonian variational ansatz (HVA) \cite{wecker_progress_2015}, which constructs the quantum circuit from the problem Hamiltonian structure, and find that a circuit depth proportional to the system size is sufficient to achieve accurate results.

As a benchmark, we apply our method to the transverse-field Ising model (TFIM) on both the one-dimensional (1D) chain and the two-dimensional (2D) square lattice, computing the 1QP dispersion in the high-field polarized phase. We demonstrate convergence of the NLCE with increasing cluster size and validate our results by comparing VQE and ED as cluster solvers. Additionally, we investigate the effect of a longitudinal field (LF) on TFIM, which breaks the parity symmetry. In this case, the ground state and 1QP excitations are no longer in different symmetry sectors, making it a particularly stringent test of our approach. The longitudinal field case requires the full machinery of PCAT to ensure cluster additivity.

Our numerical benchmarks reveal that $\lceil N/2 \rceil$ layers of HVA suffice for the pure TFIM, where NLCE+VQE results are indistinguishable from NLCE+ED. For TFIM+LF, convergence is slower but still achieved. We investigate two classes of cost functions (trace based and variance based) and find that both can achieve convergence, though with different initialization requirements and trade-offs discussed in Sec.~\ref{sec:TFIM+LF}.

The remainder of this paper is organized as follows. In Sec.~\ref{sec::overview}, we provide an overview of the NLCE+VQE approach for 1QP excitations and outline the six-step workflow. Section~\ref{sec:LCE} discusses the linked-cluster expansion for obtaining effective 1QP Hamiltonians, including details on block diagonalization (Sec.~\ref{sec:block-diag}), the NLCE method (Sec.~\ref{sec:embedding}), the details of PCAT (Sec.~\ref{sec:PCAT}), and the calculation of 1QP dispersions in the thermodynamic limit (Sec.~\ref{sec:dispersion}). Section~\ref{sec:VQE} introduces VQE as the cluster solver, with details on cost functions (Sec.~\ref{sec:vqe:cost_functions}) and state preparation (Sec.~\ref{sec:vqe:state_prep}). Section~\ref{sec:application} presents numerical results for the 1QP dispersion of the high-field phase of the TFIM, including applications to the one-dimensional chain (Sec.~\ref{sec:1D}), the two-dimensional square lattice (Sec.~\ref{sec:2D}), and the TFIM+LF (Sec.~\ref{sec:TFIM+LF}). We conclude in Sec.~\ref{sec:conclusion} with a summary and outlook.

\begin{figure*}[t]
    \centering
    \includegraphics[width=0.85\textwidth]{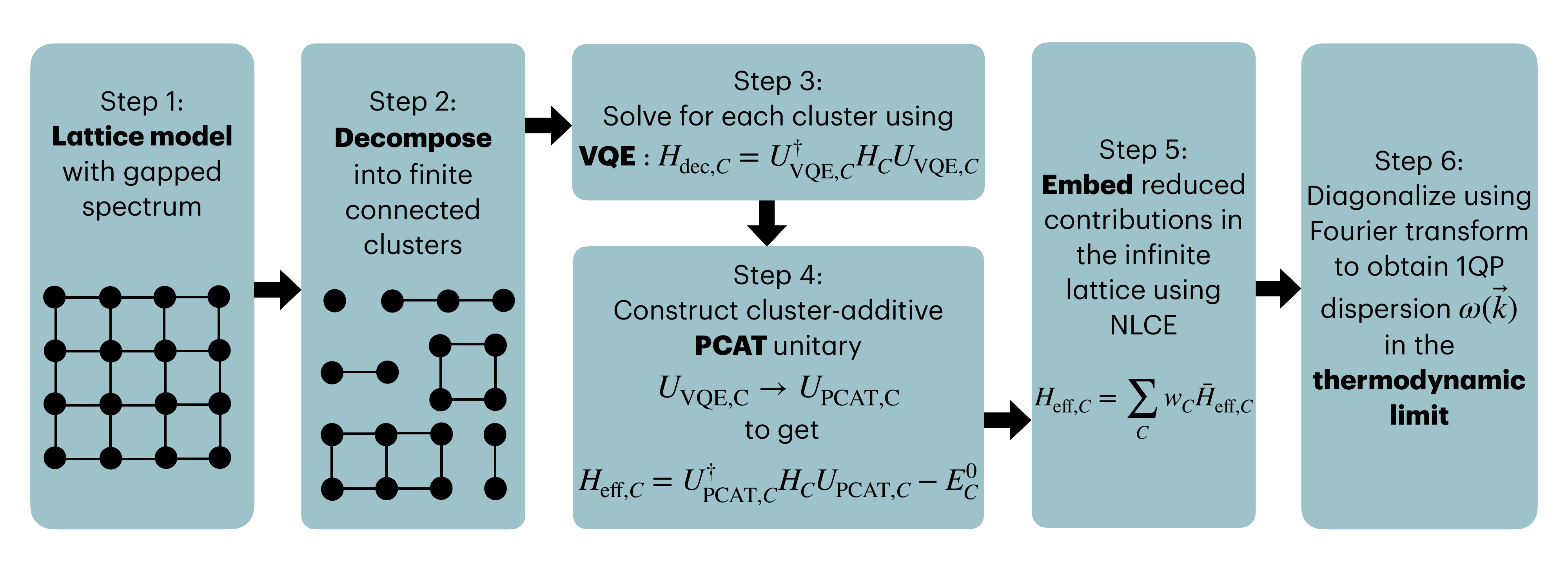}
    \caption{Workflow of the NLCE+VQE approach for computing 1QP excitation energies in the thermodynamic limit. The six-step procedure combines VQE cluster solving with PCAT postprocessing and NLCE embedding to obtain the dispersion $\omega(\vec{k})$ in the thermodynamic limit.}
    \label{fig:workflow}
\end{figure*}

\section{Overview of NLCE+VQE approach for 1QP excitations}
\label{sec::overview}

In this section, we provide an overview of our NLCE+VQE framework for computing 1QP excitations in the thermodynamic limit, outlining the key components and workflow before presenting the detailed methodology in Secs.~\ref{sec:LCE} and \ref{sec:VQE}. Here, a cluster $C$ is defined as a finite subset of a lattice consisting of a set of sites connected with bonds. Perturbative linked-cluster expansions (LCEs) are a common tool for calculating high-order series of excitation energies and spectral signatures of elementary excitations in the thermodynamic limit, but are limited by their perturbative nature. Practically, one performs perturbation theory on each cluster and then embeds the result in the thermodynamic limit by a proper embedding procedure. NLCEs replace the perturbation theory on finite clusters by numerical cluster solvers, typically ED for both scalar quantities and subspace calculations.\\

In our recent work \cite{Sumeet2024}, we introduced the NLCE+VQE approach for ground-state energies, which uses the VQE as a cluster solver to develop NLCEs for much larger cluster sizes in the future. Here, we extend this to 1QP excitations using VQE and PCAT to block diagonalize cluster Hamiltonians by decoupling the ground-state and 1QP subspace with a unitary $U_{\mathrm{PCAT}}$. Computing 1QP excitations within NLCE is substantially harder than the ground-state case. While the ground-state energy is a single scalar per cluster, the 1QP effective Hamiltonian is an $N \times N$ matrix, where $N$ is the number of sites in the cluster, so $\mathcal{O}(N^2)$ matrix elements must be computed instead of a single number. Moreover, NLCE embedding of non-scalar quantities is non-trivial since cluster additivity, which holds automatically for the ground-state energy, is not guaranteed for the 1QP block and must be enforced through PCAT. The transformed Hamiltonian is $U_{\mathrm{PCAT}}^{\dagger}HU_{\mathrm{PCAT}} = H_{\mathrm{eff}}$, the effective Hamiltonian. To obtain 1QP excitation energies, we extract the 1QP block of $H_{\mathrm{eff}}$ and subtract the ground-state energy.

We benchmark our approach on the TFIM, a paradigmatic quantum spin system that exhibits a quantum phase transition. The Hamiltonian for the TFIM with an optional longitudinal field is given by
\begin{equation}\label{eq:ham_overview}
H = - h \sum_{\nu} Z_{\nu} - J \sum_{\langle\nu^{\prime},\nu \rangle} X_{\nu^{\prime}} X_{\nu^{\phantom{\prime}}} - h_l \sum_{\nu} X_{\nu},
\end{equation}
where $Z_{\nu}$ and $X_{\nu}$ denote Pauli matrices at site $\nu$, $J$ is the nearest-neighbor Ising coupling, $h$ is the transverse-field strength, and $h_l$ is the longitudinal field. We set $h = 1$ as our energy unit throughout this work. We investigate three cases: (1) the one-dimensional TFIM chain ($h_l = 0$, Sec.~\ref{sec:1D}), (2) the two-dimensional TFIM on the square lattice ($h_l = 0$, Sec.~\ref{sec:2D}), and (3) the one-dimensional TFIM with longitudinal field ($h_l \neq 0$, Sec.~\ref{sec:TFIM+LF}). The longitudinal field breaks the $\mathbb{Z}_2$ parity symmetry, making the third case a particularly stringent test where PCAT becomes essential for ensuring cluster additivity. The expansion is developed in the high-field polarized phase ($J < h$), where the ground state is gapped and 1QP excitations correspond to localized spin flips.

We explain the workflow of our NLCE+VQE approach for 1QP energies in a stepwise fashion (see Fig.~\ref{fig:workflow}).

\begin{itemize}
    \item \textit{Step 1: Define the model. } 
    Choose a gapped quantum lattice model such as the TFIM+LF in Eq.~\eqref{eq:ham_overview}.

    \item \textit{Step 2: Decompose the lattice for NLCE.}  
    Generate topologically distinct, connected finite clusters $C$. We use rectangular graph expansions for lattice decomposition, which takes the clusters of dimensions $L_m \times L_n \leq N_{\mathrm{max}}$, as detailed in Sec.~\ref{sec:embedding}.

    \item \textit{Step 3: Solve each cluster using VQE.}
    Block diagonalize each cluster $C$ by minimizing a cost function (Sec.~\ref{sec:vqe:cost_functions}) to find optimal variational parameters $\theta$ such that $H_{\mathrm{dec}, C} = U^{\dagger}_{\mathrm{VQE}, C}(\theta)H_{C} U_{\mathrm{VQE}, C}(\theta)$ decouples the ground-state and 1QP subspace from the rest of the spectrum. 

   \item \textit{Step 4: Construct cluster-additive unitary using PCAT.} 
   Use the PCAT scheme (Sec.~\ref{sec:PCAT}) to construct a unitary $U_{\mathrm{PCAT}, C}$ from $U_{\mathrm{VQE}, C}$, yielding the effective Hamiltonian $H_{\mathrm{eff}, C} = U^{\dagger}_{\mathrm{PCAT}, C} H_{C} U^{\phantom{\dagger}}_{\mathrm{PCAT}, C}$. This transformation ensures cluster additivity: Reduced contributions of disconnected clusters vanish.

    \item \textit{Step 5: Perform NLCE embeddings.}
    Extract the 1QP effective Hamiltonian $H^{[1]}_{\mathrm{eff}, C}$ from $H_{\mathrm{eff}, C}$ and subtract the ground-state energy $E^{[0]}_C$, which is obtained from the same VQE procedure (see Sec.~\ref{sec:vqe:cost_functions}). Compute the reduced contributions $\bar{H}^{[1]}_{\mathrm{eff}, C}$ using the inclusion-exclusion principle [Eq.~\ref{eq:H_bar}] and embed into the infinite lattice, yielding $H^{[1]}_{\mathrm{eff}}$ in the thermodynamic limit.

    \item \textit{Step 6: Compute the 1QP dispersion.}  \\
     Diagonalize the effective 1QP Hamiltonian (obtained in Step 5) through Fourier transform, as described in Sec.~\ref{sec:dispersion}, to obtain the 1QP dispersion $\omega(\vec{k})$.
\end{itemize}

Finally, let us stress that PCAT is indeed a necessary ingredient of our approach. NLCEs require observables to be cluster additive, restricting calculations to connected finite clusters with disconnected parts contributing zero. For degenerate subspaces without symmetries, infinitely many unitaries can decouple the Hamiltonian, but most violate cluster additivity. The unitary $U_{\mathrm{VQE}}$ from VQE does not satisfy cluster additivity. Furthermore, unitaries must show consistency across clusters for NLCE convergence. VQE efficiently extracts the information needed for PCAT: energy expectation values and overlaps with unperturbed eigenstates from the low-energy subspace. Crucially, PCAT requires only this low-energy eigenspace information, without needing access to higher excitation sectors, avoiding exponentially costly state tomography. This suffices to construct the PCAT~\cite{hormann_projective_2023}, $U_{\mathrm{PCAT}}$, from $U_{\mathrm{VQE}}$, ensuring cluster additivity. Embedding the 1QP block of $H_{\mathrm{eff}}$ into the infinite lattice via NLCE yields the effective 1QP Hamiltonian and dispersion $\omega(\vec{k})$ in the thermodynamic limit.

The computational cost is highly nonuniform across the six steps. Steps 1, 2, 5, and 6 are classical and scale polynomially with the number of clusters, while Steps 3 and 4 carry the quantum cost. The VQE optimization in Step 3 is the dominant bottleneck. Step 4 requires a single round of $\mathcal{O}(N^2)$ measurements on the converged circuit to extract the matrices needed for PCAT, and avoids the exponential cost of full state tomography.\\

\section{Linked-cluster expansion for effective one-quasiparticle picture}
\label{sec:LCE}

In this section, we establish the mathematical and computational framework for extracting 1QP excitations in the thermodynamic limit using NLCEs. We first introduce the concept of block diagonalization and explain why cluster additivity is essential for NLCE convergence. We then present the NLCE formalism for 1QP excitations and describe how the PCAT ensures that the effective Hamiltonian satisfies the required cluster-additivity property. Finally, we show how the 1QP dispersion relation is obtained from the effective Hamiltonian via Fourier transformation.

\subsection{Block diagonalization}\label{sec:block-diag}

A quasiparticle ansatz typically characterizes the low-energy excitation spectrum by non-interacting dressed excitations. These correspond to block diagonalizing a strongly correlated many-body Hamiltonian into independent sectors labeled by quasiparticle number. Formally, given a Hamiltonian
\begin{equation}
  H = H_{0} + xV,
\label{eq:ham_decomp}
\end{equation}
with unperturbed part $H_{0}$ and perturbation $V$ at strength $x$, we seek to partition the Hilbert space into subspaces based on the number of elementary excitations above the ground state. The unperturbed Hamiltonian $H_{0}$ is assumed to be exactly solvable and serves as the reference. Its eigenstates organize the Hilbert space into quasiparticle sectors. Denoting by $E_{0}^{[n]}$ the unperturbed energy of the $n$ quasiparticle subspace, the Hilbert space decomposes as
\begin{equation}
  \mathcal{H}=\bigoplus_n\mathcal{H}_{0}^{[n]}.
  \label{eq:H_space_decomp}
\end{equation}

Each $\mathcal{H}_{0}^{[n]}$ carries all states with eigenvalue $E_{0}^{[n]}$ of $H_{0}$ corresponding to $n$ quasiparticles. We call $H_{0}$ block diagonal if
\begin{equation}
  H_{0}=\bigoplus_{n=0}^N H_{0}^{[nn]}
  \label{eq:block_diag_form}
\end{equation}
and $H_{0}^{[nn]}$ keeps $\mathcal{H}_{0}^{[n]}$ invariant. We call a transformation $U$ block diagonalizing if
\begin{equation}
  H_{\mathrm{eff}}=U^{\dagger}HU,
  \label{eq:H_eff}
\end{equation}
and $H_{\mathrm{eff}}$ retains the block structure of $H_{0}$, i.e.,

\begin{equation}
  H_{\mathrm{eff}}=\bigoplus_{n=0}^N H_{\mathrm{eff}}^{[nn]}.
\end{equation}

Here, $H_{\mathrm{eff}}^{[nn]}$ denotes the block corresponding to the $n$ quasiparticle sector. For example, $H_{\mathrm{eff}}^{[11]}$ is the $N \times N$ matrix for the 1QP block, where $N$ is the number of sites.

The subspace $\mathcal{H}_{\mathrm{eff}}^{{[n]}}$ is defined as containing those eigenstates of $H$ that adiabatically follow the states of $\mathcal{H}_{0}^{[n]}$ as the perturbation strength $x'$ varies from 0 to $x$ in $H = H_{0} + x'V$.

\begin{figure}[t]
  \centering
  \includegraphics[width=0.6\linewidth]{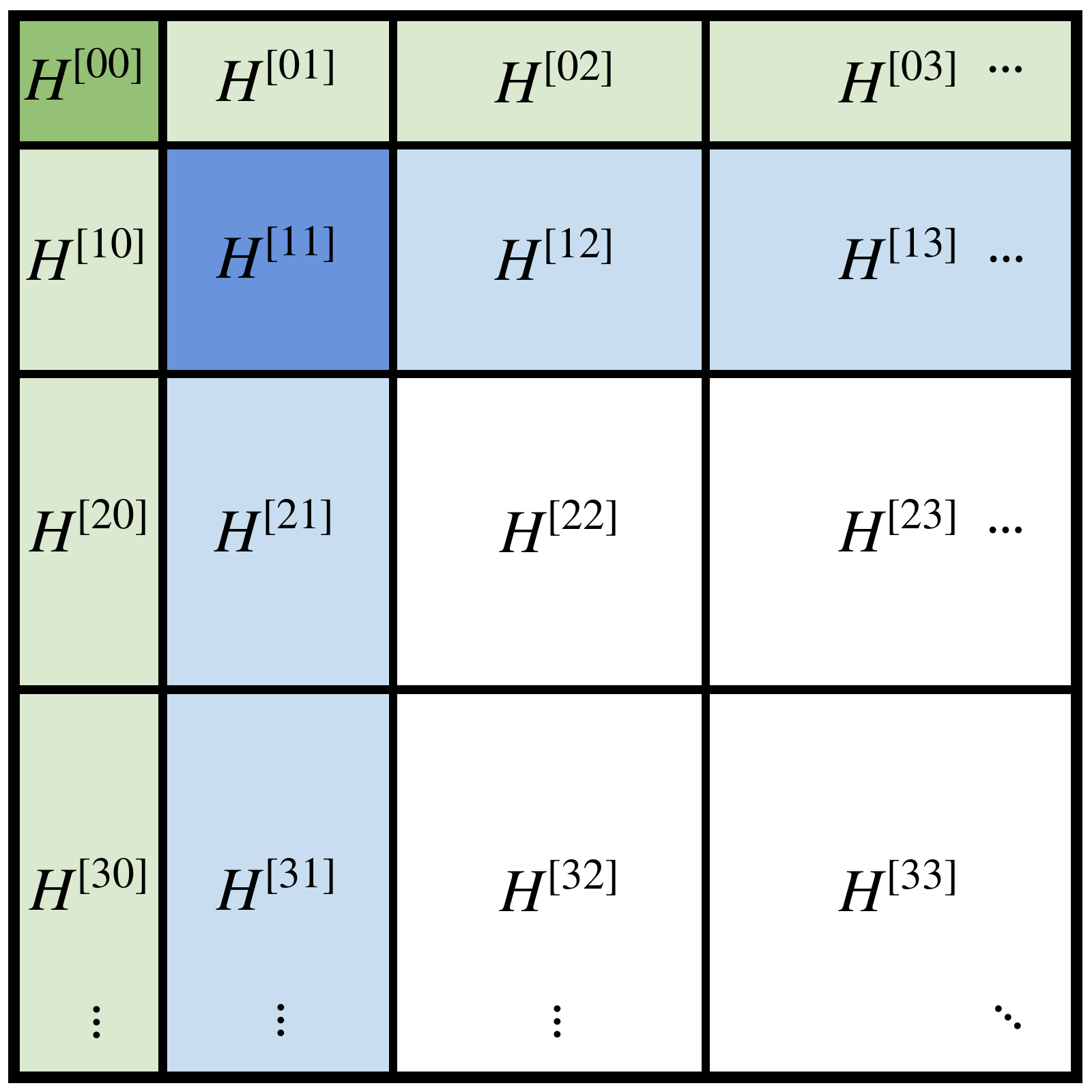}
  \caption{Diagrammatic view of a block-diagonal Hamiltonian. The ground-state ($0$QP) block is shown in dark green and the $1$QP block is shown in dark blue. The light shaded blocks are off-diagonal terms that are minimized during block diagonalization.}
  \label{fig:ham_block}
\end{figure}

In realistic settings, one can not block diagonalize the full Hamiltonian but only the low-energy part. In our case, we restrict the decomposition to three sectors consisting of the ground-state subspace $\mathcal{H}_{\mathrm{eff}}^{[0]}$, the 1QP subspace $\mathcal{H}_{\mathrm{eff}}^{[1]}$, and all higher excitations.
The pictorial representation of such a block-diagonal Hamiltonian is shown in Fig.~\ref{fig:ham_block}, where the block diagonalizing $U$ should eliminate the faint blue and green off-diagonal sectors $H_{\mathrm{eff}}^{[nm]}$ with $n \neq m$.

A complete diagonalization of an infinite system is intractable, but LCEs provide a systematic approximation by combining results from finite clusters to recover thermodynamic-limit properties. For LCEs to converge, the transformation must be cluster additive: Disconnected clusters should contribute zero. However, even for the block-diagonalizing part of $U$ infinitely many choices exist for a degenerate subspace $n$ as discussed in Ref.~\cite{Cederbaum1989}. While the ground state is uniquely determined, any unitary rotation within the 1QP subspace yields an equally valid block diagonalization due to the quasi-degeneracy. Higher-energy sectors have even more freedom.

This freedom necessitates, but also allows one to tailor the structure of the effective Hamiltonian to ensure cluster additivity. In particular, the PCAT~\cite{hormann_projective_2023} yields a specific unitary $U_{\mathrm{PCAT}}$ that preserves cluster additivity. Cluster additivity ensures that only connected clusters need to be calculated, as disconnected clusters contribute zero, significantly reducing computational cost.

\subsection{Linked-cluster expansions}\label{sec:embedding}
In this subsection, we first outline the LCE framework assuming a suitable effective Hamiltonian exists, and then specify the requirements it must satisfy.

The general formalism for perturbative and nonperturbative LCEs is reviewed in Refs.\,\cite{Oitmaa2006,TANG2013557,Rigol2006}. 
First introduced for perturbative calculations of ground-state energies in quantum many-body systems, LCEs were subsequently extended to the study of excitation energies \cite{Gelfand1996}. The core idea of LCEs is to decompose the lattice into topologically distinct, finite connected subgraphs called clusters $C$, and use the inclusion-exclusion principle to sum up their additive contributions to recover thermodynamic-limit properties.    Let us for the moment assume that the 1QP effective Hamiltonian $H^{[1]}_{\mathrm{eff}, C}$ is such an additive quantity acting as a linear operator in the space of $N$ excitations. Then, the LCE scheme goes as follows.

A cluster is a finite set of sites connected by bonds, where sites carry on-site Hamiltonian terms (transverse and longitudinal fields) and bonds carry the two-body interactions ($XX$ couplings). These clusters are chosen with rectangular graph expansion having the dimensions $L_m \times L_n \leq N_{\mathrm{max}}$ where $N_{\mathrm{max}}$, is the maximum number of spins taken in NLCE calculation.\\

\noindent The reduced contribution of a cluster $C$ is defined recursively as
\begin{equation}
    \bar{H}^{[1]}_{\mathrm{eff}, C} = H^{[1]}_{\mathrm{eff}, C} - \sum_{C'\subsetneq C} w_{C'/C}\, \bar{H}^{[1]}_{\mathrm{eff}, C'},
  \label{eq:H_bar}
\end{equation}
where $w_{C'/C}$ counts embeddings of the subgraph $C'$ inside $C$. 
Summing the reduced contributions gives the effective 1QP Hamiltonian in the thermodynamic limit, 
\begin{equation}
  H^{[1]}_{\mathrm{eff}} = \sum_{C} w_{C}\, \bar{H}^{[1]}_{\mathrm{eff}, C},
\end{equation}
where weight $w_{C}$ counts embeddings of cluster $C$ per lattice site. 

\begin{figure}
    \centering
    \includegraphics[width=1\linewidth]{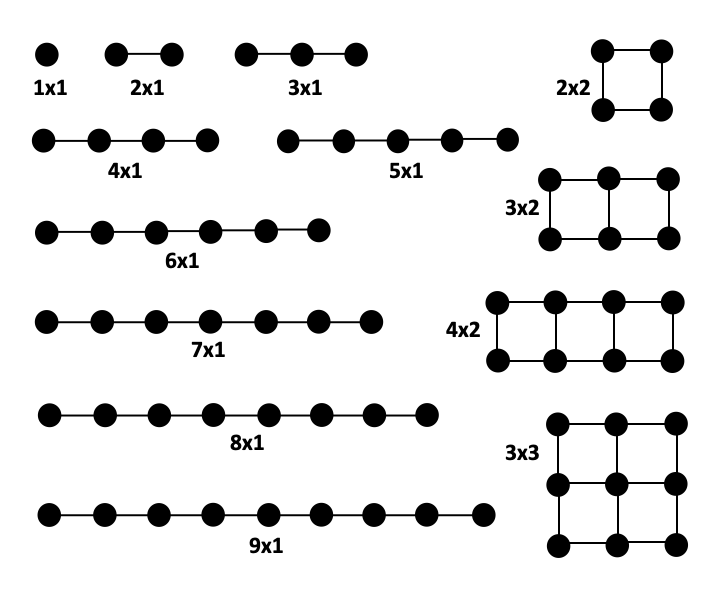}
    \caption{Rectangular graph expansion for the square lattice, up to $N_\mathrm{max} = 9$, showing all clusters of dimensions  $L_m \times L_n$ with $L_m \cdot L_n \leq N_{\mathrm{max}}$. For the NLCE embedding, each non-square cluster ($L_m \neq L_n$) carries weight $w_C = 2$ from its two lattice orientations (horizontal and vertical), while the $2 \times 2$ and $3 \times 3$ square cluster have $w_C = 1$.}
    \label{fig:rect_graphs}
\end{figure}

Often, one uses a full graph decomposition for the NLCE where the calculation is performed on topologically distinct graphs of minimal size. However, the number of graphs increases exponentially with the number of sites, which is challenging for the NLCE+VQE approach. As for the NLCE+VQE for ground-state energy \cite{Sumeet2024}, we adopt a rectangular graph expansion \cite{Enting1996} for NLCE. In this expansion, only clusters of size $L_m \times L_n$ with $L_m \cdot L_n \leq N_{\mathrm{max}}$ are included, where $N_{\mathrm{max}}$ is the maximum number of spins in the largest cluster of the NLCE. These clusters have open boundary conditions. The rectangular graph expansion is applicable for the chain and the square lattice discussed below and is beneficial because the number of clusters increases polynomially with $N$. For the one-dimensional chain, this reduces to considering open segments of length $L_m \leq N_{\mathrm{max}}$. We will apply this expansion to the TFIM+LF in Sec.~\ref{sec:application}. The rectangular graph expansion for the square lattice is illustrated in Fig.~\ref{fig:rect_graphs}.

We now specify the calculation of $H^{[1]}_{\mathrm{eff}, C}$ and how it can be made into a cluster-additive quantity. The effective 1QP Hamiltonian on a cluster $C$ with $N$ 1QP states is an $N\times N$ matrix defined as
\begin{equation}
    H^{[1]}_{\mathrm{eff}, C} = (U^\dagger H U)^{[11]} - E^{[0]}_{C},
\end{equation}

where $^{[11]}$ denotes the 1QP block and $U$ is the block-diagonalizing transformation. $E^{[0]}_C$ is obtained from the same VQE procedure: either via a separate ground-state minimization (separate-unitary case), or in the single-unitary case via diagonalization of the Hamiltonian projected onto the VQE-prepared subspace (see Sec.~\ref{sec:hardware}). The ground-state energy $E^{[0]}_{C}$ must be subtracted to ensure that $H^{[1]}_{\mathrm{eff}, C}$ is intensive \cite{Gelfand1996}. Importantly, since $E^{[0]}_C$ enters only as a diagonal shift of $H^{[1]}_{\mathrm{eff},C}$, any estimation error from VQE affects only the on-site energies and leaves the hopping matrix elements unchanged. After Fourier transformation in the NLCE embedding, this propagates at most to a momentum-independent shift of $\omega(\vec{k})$, leaving its shape and bandwidth unaffected. 

The need for subtracting the ground-state energy is best understood through a concrete example. Consider two disconnected clusters $A$ and $B$ (i.e., they share no sites or bonds). Since the Hamiltonians $H_{A}$ and $H_{B}$ on the two disconnected clusters commute, $[H_{A}\otimes\mathds{1}_{B},\mathds{1}_{A}\otimes H_{B}]=0$, it follows for the eigenfunctions on $A \cup B$ that

\begin{equation}
  \ket{\Psi^{[n+m]}_{A\cup B}}\;=\;\ket{\Psi^{[n]}_{A}} \otimes \ket{\Psi^{[m]}_{B}}.
  \label{eq:psi_AuB}
\end{equation}

Here, $|\Psi^{[n]}\rangle$ denotes an eigenstate of the full Hamiltonian $H$ with $n$ quasiparticles, and $[n]$ ($[m]$) refers to the number of quasiparticles on cluster $A$ ($B$). The effective Hamiltonian part on $A\cup B$ that describes 1QP excitations within $B$ would always have a contribution of the ground-state energy on $A$. Hence, one has to subtract it ($n=0$ and $m=1$ in the above formula).

This is linked to the concept of cluster additivity. A transformation $U$ is cluster-additive if 
\begin{equation}
    U^\dagger H_{A\cup B}U = H_{\mathrm{eff}, A\cup B} = H_{\mathrm{eff}, A}\otimes\mathds{1}_{B} + \mathds{1}_{A}\otimes H_{\mathrm{eff}, B}.
\end{equation}
If fulfilled, 
\begin{equation}
  \bar{H}^{[1]}_{\mathrm{eff}, A\cup B}
  \;=\;
  H^{[1]}_{\mathrm{eff}, A\cup B} - E^{[0]}_{A\cup B}
  \;=\;
  \bar{H}^{[1]}_{\mathrm{eff}, A}
  \oplus
  \bar{H}^{[1]}_{\mathrm{eff},B}
 \end{equation}
will hold. Subtracting the ground-state energy alone is necessary, but not sufficient to ensure cluster additivity.

In the next subsection, we explain how PCAT constructs a transformation $U_{\mathrm{PCAT}}$ that guarantees both properties using only limited information from the ground-state and 1QP subspaces.

\subsection{Projective cluster-additive transformation}\label{sec:PCAT}

In LCEs, both perturbative and nonperturbative, it is essential that observables on large systems can be consistently constructed from smaller, disconnected clusters. To understand why PCAT \cite{hormann_projective_2023} achieves cluster additivity, we first establish the mathematical requirements. 

Two clusters are disconnected if the unperturbed Hamiltonian is cluster additive on them (see Sec.~\ref{sec:embedding}) \cite{knetter2003structure}:
\begin{equation}
  H_{A\cup B}\;=\;H_{A}\otimes\mathds{1}_{B} + \mathds{1}_{A}\otimes H_{B}.
  \label{eq:H_AuB}
\end{equation}
We demand of a transformation $U$ that when applied to the full Hamiltonian \mbox{$H \;=\; H_{0} + x V$}, $U^\dagger H U^{\phantom \dagger}$ remains cluster additive on two disconnected clusters $A$ and $B$. This ensures that no nonlinked clusters contribute to the expansion. We first illustrate why standard approaches fail to maintain cluster additivity, then show how PCAT resolves this issue through a modified state construction.

To understand that PCAT fulfills cluster additivity, we have to investigate the product state structure of eigenstates on $A \cup B$. Let $\{\ket{\Phi^{[n]}_{i}}\}$ denote the eigenstates of $H_{0}$ in the $(n+1)$th subspace, and let $\{\ket{\Psi^{[n]}_{i}}\}$ denote the corresponding eigenstates of the full Hamiltonian $H$, where $i$ labels states within the degenerate manifold. $P^{[n]}$ is the corresponding projector, and $\bar P^{[n]}$ the adiabatically connected projector of the full Hamiltonian $H$. On $A \cup B$ eigenstates have the product form as in Eq.~\ref{eq:psi_AuB}.
% \begin{equation}
%   \ket{\Psi^{[n+m]}_{A\cup B}}\;=\;\ket{\Psi^{[n]}_{A}} \otimes \ket{\Psi^{[m]}_{B}}.
%   \label{eq:psi_AuB}
% \end{equation}

It is instructive to see why a widely used approach to decouple an energy block fails to be cluster additive: The canonical two-block (2b) transformation (Schrieffer-Wolff, des Cloizeaux, Takahashi, etc. \cite{Oitmaa2006, DESCLOIZEAUX1960321, Cederbaum1989, Takahashi, Shavitt1980, Schrieffer-Wolff}), which can be written in projector form for the transformation restricted to block $n$ as
\begin{equation}
      U^{[n]}_{\mathrm{2b}} = \bar{P}^{[n]} (P^{[n]} \bar{P}^{[n]} P^{[n]})^{-1/2},
\end{equation}
where $(M)^{-1/2}$ for a rank-deficient matrix $M$ denotes the inverse square root restricted to the range of $M$, or equivalently, the Moore-Penrose pseudoinverse square root. This convention applies throughout this section whenever inverse operations are applied to projector products.
The full transformation $U_{\mathrm{2b}} = \sum_n U_{\mathrm{2b}}^{[n]}$ minimizes $\|1 - U_{\mathrm{2b}}\|$  . Let $U^{[n]}$ denote the $2^N \times N_n$ matrix whose columns are the exact eigenvectors of $H$ corresponding to the $n$th subspace, where $N_n$ is the dimension of that subspace. When embedded in the full $2^N \times 2^N$ space (as needed for sums like $\sum_n U^{[n]}$), this is understood as a partial isometry extended by zeros outside the $n$-quasiparticle subspace. The transformation can also be given in terms of the overlap matrix $X^{[n]}$, an $N_n \times N_n$ matrix with elements\\

\begin{equation}
X^{[n]}_{ij} = \langle \Phi^{[n]}_i | \Psi^{[n]}_j \rangle,
\end{equation}
where $|\Phi^{[n]}_i\rangle$ are the unperturbed eigenstates and $|\Psi^{[n]}_j\rangle$ are the eigenstates of $H$ in block $n$. Then,
\begin{equation}
    U_{2b}^{[n]} = U^{[n]} X^{[n] \dagger} (X^{[n]} X^{[n] \dagger})^{-1/2}.
\end{equation}

Written this way, the problem with cluster additivity becomes apparent: For simplicity, consider the 1QP excitations ($n=1$). Then $X^{[1]}$ becomes non-block-diagonal for disconnected clusters owing to the product form when the unperturbed ground state overlaps with excited states (i.e., $P^{[1]} \ket{\Psi^{[0]}} \neq 0$). Contributions from $\ket{\Psi^{[0]}_{A}} \otimes \ket{\Psi^{[1]}_{B}}$ lead to nonzero elements in $X^{[1]}$ at the same entries as $\ket{\Psi^{[1]}_{A}} \otimes \ket{\Psi^{[0]}_{B}}$. This leads to intercluster matrix elements between blocks $A$ and $B$. It also modifies matrix elements in the $A$-only or $B$-only blocks compared to the isolated $A$ and $B$ clusters.

\begin{figure}
    \centering
    \includegraphics[width=0.8\linewidth]{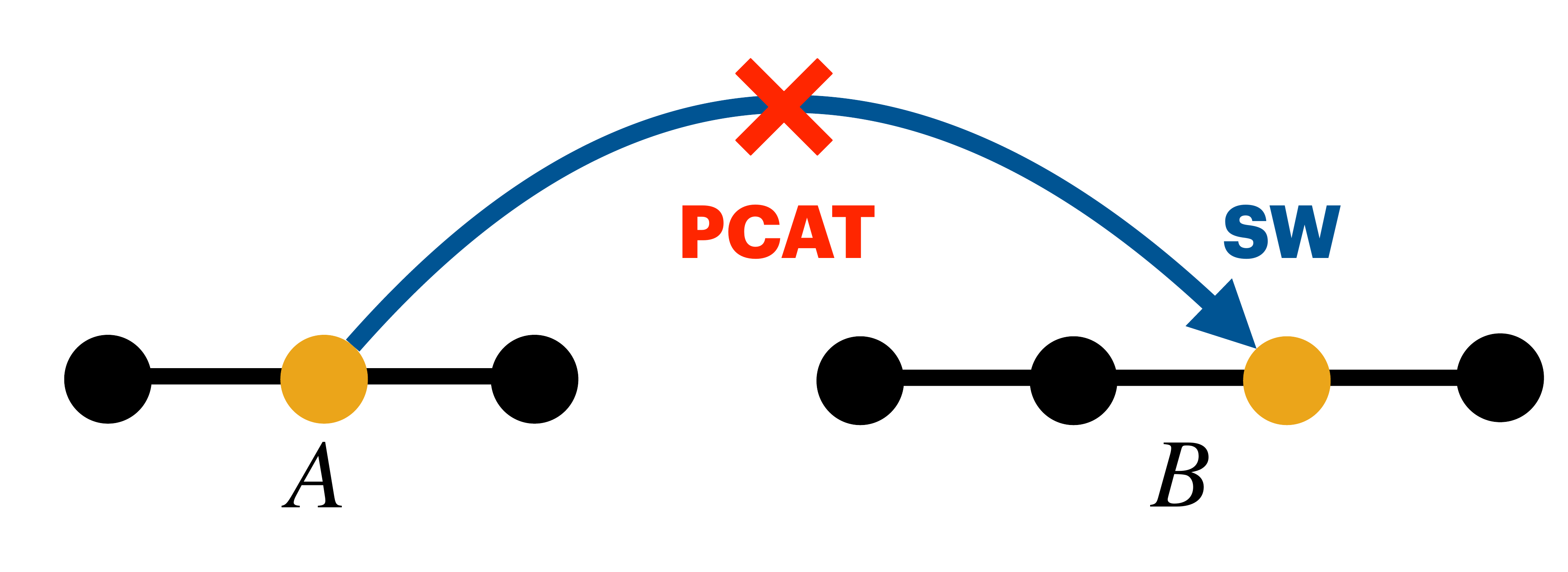}
    \caption{Comparison of 2b and PCAT transformations on disconnected clusters. The 2b transformation allows unphysical hopping of a dressed quasiparticle (yellow) between clusters $A$ and $B$ that share no bonds. PCAT forbids such hopping, ensuring cluster additivity.}
    \label{fig:hopping}
\end{figure}

Thus, cluster additivity is violated and dressed quasiparticles can hop between disconnected clusters (see Fig.~\ref{fig:hopping}). In the thermodynamic limit, disconnected clusters can be arbitrarily far apart, so such hopping represents unphysical infinite-range coupling. This artifact arises because the 2b transformation creates entanglement between clusters that should be independent.

In order to resolve this, PCAT \cite{hormann_projective_2023} constructs modified states $\{\,|\widetilde{\Psi}^{[n]}\rangle\}$ from the original eigenstates by eliminating all components projected onto lower-energy subspaces of $H_{0}$, using only subtractions of states from the low-energy spectrum. These modified states lie within the span of low-energy eigenspaces but are not eigenstates themselves. From these states, one constructs a modified overlap $\widetilde{X}^{[n]}$, which in contrast to $X^{[n]}$ is now block diagonal again on $A$ and $B$. In detail, the construction of modified states goes as follows.

The recursion for constructing modified 1QP states is
\begin{align}
  &|\widetilde{\Psi}^{[0]}\rangle = |\Psi^{[0]}\rangle, \nonumber\\[2pt]
  &|\widetilde{\Psi}^{[1]}_{i}\rangle =
     |\Psi^{[1]}_{i}\rangle -
     \frac{\langle\Phi^{[0]}|\Psi^{[1]}_{i}\rangle}
          {\langle\Phi^{[0]}|\Psi^{[0]}\rangle}|\Psi^{[0]}\rangle,
     \label{eq:pcat-recursion-1qp}
\end{align}
which ensures that $P^{[0]}|\widetilde{\Psi}^{[1]}_{i}\rangle=0$ for all 1QP states. This removes any admixture of the ground state from the 1QP eigenstates. Note that in the special case where $\langle\Phi^{[0]}|\Psi^{[1]}\rangle = 0$ the transformation remains the same for the one-particle subspace. This is the case for the transverse-field Ising model, but as soon as the parity is broken, e.g., by a longitudinal field, the corrections become necessary.

For higher subspaces, the construction generalizes by recursively projecting out all unperturbed lower-energy subspaces. Using the low-energy projector
\begin{equation}
  R^{[n]} = \sum_{m<n} P^{[m]}
\end{equation}
and its adiabatically connected counterpart $\bar{R}^{[n]}$, the general subtraction scheme reads
\begin{equation}
  |\widetilde{\Psi}^{[n]}\rangle =
     \left(1 - \bar{R}^{[n]} (R^{[n]} \bar{R}^{[n]} R^{[n]})^{-1} R^{[n]}\right)
     |\Psi^{[n]}\rangle.
\end{equation}
This ensures that each modified state $|\widetilde{\Psi}^{[n]}\rangle$ has no projection onto lower-energy subspaces of $H_{0}$.

The key insight is that a state $|\Psi^{[n+m]}_{A\cup B}\rangle = |\Psi^{[n]}_{A}\rangle \otimes |\Psi^{[m]}_{B}\rangle$ is transformed into
\begin{equation}
  |\widetilde{\Psi}^{[n+m]}_{A\cup B}\rangle = |\widetilde{\Psi}^{[n]}_{A}\rangle \otimes |\widetilde{\Psi}^{[m]}_{B}\rangle.
\end{equation}
This factorization is the essence of why PCAT preserves cluster additivity. Since modified states on $A\cup B$ factor into modified states on $A$ and $B$ separately, the transformation introduces no artificial correlations between disconnected clusters. Each cluster's quasiparticles remain localized on the cluster.

From this factorization property, several important consequences follow. In particular, it yields $P^{[s]}|\widetilde{\Psi}^{[n+m]}_{A\cup B}\rangle = 0$ for $s < n + m$ but also $P^{[s]}|\widetilde{\Psi}^{[n]}_{A}\rangle = 0$ for $s < n$ and $P^{[s]}|\widetilde{\Psi}^{[m]}_{B}\rangle = 0$ for $s < m$. Indeed, the identity for modified states on disconnected clusters $A$ and $B$ is sufficient to guarantee cluster additivity of the transformation:
\begin{equation}
   U_{\mathrm{PCAT}}^{\dagger}
   H_{A\cup B}\,
   U_{\mathrm{PCAT}}^{\phantom{\dagger}}
   = H_{\mathrm{eff}, A}\otimes\mathds{1}_{B}
     +\mathds{1}_{A}\otimes H_{\mathrm{eff}, B}.
\end{equation}

We can now write the explicit form of the $U_{\mathrm{PCAT}}$ transformation. The modified overlap matrix $\widetilde{X}^{[n]}$ is an $N_n \times N_n$ matrix with elements
\begin{equation}
\widetilde{X}^{[n]}_{ij} = \langle \Psi^{[n]}_i | \widetilde{\Psi}^{[n]}_j \rangle,
\end{equation}
where $|\Psi^{[n]}_i\rangle$ are the unperturbed eigenstates and $|\widetilde{\Psi}^{[n]}_j\rangle$ are the modified eigenstates in block $n$. The transformation for a single block $n$ is then
\begin{equation}
  U_{\mathrm{PCAT}}^{[n]} = U^{[n]} \widetilde{X}^{[n]\dagger} (\widetilde{X}^{[n]} \widetilde{X}^{[n]\dagger})^{-1/2},
\end{equation}
and for the full Hilbert space (with each block extended to $2^N \times 2^N$ as partial isometries),
\begin{equation}
  U_{\mathrm{PCAT}} = \sum_n U_{\mathrm{PCAT}}^{[n]}.
\end{equation}

Since the modified states $|\widetilde{\Psi}^{[n]}\rangle$ can be written using only projectors (as in the $R^{[n]}$ form above), one can write $U_{\mathrm{PCAT}}$ in terms of projectors only, which shows that it has a unique local perturbative expansion when using Kato's perturbative series \cite{Kato1949} for projection operators. This property implies that the transformation has the same form on every cluster. It enables a consistent perturbative expansion across clusters and ensures that a given perturbative order can be reached using the smallest possible clusters. 

The projector form also establishes that the transformation satisfies
\begin{equation}
U_{\mathrm{PCAT}}^{[n]}(U^{[0]},U^{[1]},\dots,U^{[n]}) = U_{\mathrm{PCAT}}^{[n]}(U^{[0]},\dots,U^{[n]} \mathcal{W})
\end{equation}
for arbitrary unitaries $\mathcal{W} \in \mathcal{U}(N_n)$ (the group of $N_n \times N_n$ unitary matrices) acting within subspace $n$, as rotations within a subspace do not affect the projection operators. This invariance property is essential for practical implementations.

Thus, PCAT yields a local, cluster-additive effective Hamiltonian built only from the low-energy eigenspaces and is compatible with both perturbative and nonperturbative LCEs.

This highlights the global character of the transformation; i.e., it can be fully constructed from the eigenspace of interest and those of lower energy. This global character is essential for our approach: VQE (Sec.~\ref{sec:VQE}) targets only the low-energy eigenspace, and PCAT constructs the transformation from this limited information, namely, energy expectations and state overlaps, without requiring knowledge of higher excitation sectors.

\subsection{1QP dispersion in the thermodynamic limit}\label{sec:dispersion}

Once we obtain the effective Hamiltonian for finite clusters, the goal is to obtain the effective 1QP Hamiltonian and therefore the 1QP dispersion in the thermodynamic limit. For translationally invariant systems, 1QP eigenstates are labeled by momentum $\vec{k}$. With lattice spacing set to unity and assuming a single-site unit cell, the Brillouin zone is $k \in [-\pi, \pi]$ for the one-dimensional chain and $\vec{k} = (k_x, k_y) \in [-\pi, \pi]^2$ for the square lattice.
 
For this purpose, the rectangular clusters are embedded into the infinite lattice as explained in Sec.~\ref{sec:embedding} using the inclusion-exclusion principle. For the rectangular expansion, the embedding weight $w_{C}$ for a cluster of size $L_m\times L_n$ on an infinite lattice is simply unity (one embedding per lattice site in the thermodynamic limit). For translationally invariant models, eigenstates are also momentum eigenstates and Fourier transforming the real-space Hamiltonian gives the 1QP energies. The dispersion relation for both the chain and square lattice reads
\begin{equation}
    \omega(\vec{k}) = \sum_{L_m,L_n} \sum_{\vec{\mu},\vec{\nu}} \exp \big( i \vec{k} \cdot (\vec{\nu}-\vec{\mu})\big) \bar{H}^{[1]}_{\mathrm{eff}, L_m \times L_n,\vec{\mu},\vec{\nu}}\, ,
\end{equation}

where $\vec{\mu}, \vec{\nu}$ label lattice sites within the cluster. For the one-dimensional chain, $L_n = 1$ and $\vec{k}$ reduces to scalar momentum $k$. For the square lattice, $\vec{k} = (k_x, k_y)$ spans the two-dimensional Brillouin zone.\\
The calculation of reduced contribution is particularly simple for the rectangular expansion. It is easy to see that the contribution $\bar{H}^{[1]}_{\mathrm{eff}, L_m \times L_n,\vec{k}}$ of a cluster in $k$ space can be rewritten in terms of non-reduced contributions in the following way:
\begin{equation}\label{eqn:rect_exp}
\begin{aligned}
    \bar{H}^{[1]}_{\mathrm{eff},L_m \times L_n,\vec{k}} & = H^{[1]}_{\mathrm{eff}, L_m \times L_n,\vec{k}}  \\ 
    & - 2 H^{[1]}_{\mathrm{eff}, L_m-1 \times L_n,\vec{k}} - 2 H^{[1]}_{\mathrm{eff}, L_m \times L_n-1,\vec{k}} \\ 
    & + 4 H^{[1]}_{\mathrm{eff}, L_m-1 \times L_n-1,\vec{k}}\\
    & +  H^{[1]}_{\mathrm{eff}, L_m-2 \times L_n,\vec{k}}  +  H^{[1]}_{\mathrm{eff}, L_m \times L_n-2,\vec{k}} \\ 
    & - 2 H^{[1]}_{\mathrm{eff}, L_m-1 \times L_n-2,\vec{k}} - 2 H^{[1]}_{\mathrm{eff}, L_m-2 \times L_n-1,\vec{k}}  \\ 
    & + H^{[1]}_{\mathrm{eff}, L_m-2 \times L_n-2,\vec{k}}\, . 
    \end{aligned}
\end{equation}
The coefficients in this expression and the alternating signs follow from the inclusion-exclusion principle applied to rectangular geometry. 

Each subtracted term removes overcounting from smaller clusters. This identity reveals an important insight: Only the largest clusters and their boundaries contribute to the rectangular expansion, as interior contributions cancel in the inclusion-exclusion sum. This property reduces error accumulation. Specifically, once clusters of size $L_m \times L_n$ are included in the expansion, all clusters of size $(L_m-2) \times (L_n-2)$ or smaller have reduced contributions that vanish identically. Their physics is therefore fully captured by the larger embeddings. This is analogous to the 1D case where only clusters of size $N$ and $N-1$ contribute to the dispersion at order $N_{\mathrm{max}} = N$.

Having established the NLCE framework, we now describe how VQE is used as the cluster solver.

\section{VQE as cluster solver}
\label{sec:VQE}

The VQE is a hybrid quantum-classical algorithm for near-term quantum devices. In this section, we describe how VQE serves as the cluster solver within our NLCE framework, block-diagonalizing the Hamiltonian on each finite cluster to decouple the ground state and 1QP subspace from higher excitations.

As discussed in Sec.~\ref{sec:PCAT}, the unitary transformation $U_{\mathrm{VQE}}$ obtained from VQE optimization alone is generally not cluster additive and therefore not suitable for direct use in NLCE. Instead, VQE provides the information, namely, the low-energy eigenspaces and their overlaps with the unperturbed states, needed to construct the cluster-additive transformation $U_{\mathrm{PCAT}}$ via the PCAT scheme. From the full effective Hamiltonian $H_{\mathrm{eff}} = U_{\mathrm{PCAT}}^\dagger\,H\,U_{\mathrm{PCAT}}$, we extract the 1QP block
\begin{equation}
H_{\mathrm{eff}}^{[1]} = (U_{\mathrm{PCAT}}^\dagger\,H\,U_{\mathrm{PCAT}})^{[11]} - E^{[0]},
\end{equation}
where the superscript $^{[11]}$ denotes the 1QP block and $E^{[0]}$ is the ground-state energy.

A key advantage of the PCAT approach is that it requires only energy expectation values and overlaps, quantities that VQE can provide, rather than full state tomography, which would be exponentially costly. This efficiency stems from the structure of PCAT itself, as detailed in Sec.~\ref{sec:PCAT}.

Throughout the paper, we use VQE as cluster solver, though the PCAT framework is algorithm agnostic and applies equally to other eigenstate preparation methods such as quantum phase estimation or adiabatic state preparation.

VQE operates by minimizing a cost function $C(\theta)$ that depends on variational parameters $\theta$ characterizing a quantum circuit $U_{\mathrm{VQE}}(\theta)$. For ground-state calculations, the cost function is simply the energy expectation value, and the Rayleigh-Ritz variational principle ensures convergence to the lowest-energy state. For excited states, the situation is more subtle. Several approaches exist within the VQE framework for accessing excited states, including deflation methods \cite{Higgott2019}, purified state approaches \cite{Benavides-Riveros2022}, and subspace-search methods \cite{Nakanishi2019}. Deflation methods sequentially compute excited states by enforcing orthogonality with previously found states. Purified state and subspace-search methods treat entire degenerate subspaces simultaneously, which is more natural for our purpose where we need to decouple a full quasidegenerate 1QP subspace on finite clusters.

The performance of VQE depends on two components: (1) the cost function, which must be designed to achieve block diagonalization rather than simply finding individual eigenstates, and (2) the ansatz for the variational circuit $U_{\mathrm{VQE}}(\theta)$, which must be sufficiently expressive yet trainable. For the ansatz, we adopt the HVA, which we successfully employed for ground-state energy calculations in our previous NLCE+VQE work \cite{Sumeet2024}. 
% We note that particle-number-preserving ans\"atze are not suitable in our setting: the models we consider do not conserve the bare quasiparticle number, so 
% A fixed-particle-number ansatz would remain confined to one sector and could not represent these states.
 The HVA is also well suited for excited states. Since the eigenstates of the full Hamiltonian are superpositions across multiple bare quasiparticle sectors of $H_0$, the ansatz must generate inter-sector mixing. The HVA achieves this naturally, as it is constructed directly from the terms of $H$.

The remainder of this section is organized as follows. In Sec.~\ref{sec:vqe:cost_functions}, we discuss the design of cost functions for block diagonalization, introducing both variance-based and trace-based approaches and comparing their properties. In Sec.~\ref{sec:vqe:state_prep}, we describe the HVA ansatz and our initialization strategy. Finally, in Sec.~\ref{sec:vqe:measurements}, we detail the quantum measurement protocol required to construct PCAT. This subsection describes how to extract energy expectations and state overlaps from the VQE solution and presents the explicit algorithm for computing $U_{\mathrm{PCAT}}$.

\subsection{Cost functions}
\label{sec:vqe:cost_functions}
For clarity, all cost functions below are evaluated on a finite cluster $C$ within the NLCE, though we suppress the cluster label for notational simplicity. The unitary $U$ appearing in the cost functions defined in this section is generic and can be realized by any ansatz.

VQE aims to find the parameters $\theta$ that minimize a cost function $C(\theta)$. For ground states, the ground-state energy $E^{[0]}_C$ can be obtained separately by direct energy minimization via the Rayleigh-Ritz principle,
\begin{equation}
    C^{\mathrm{GS}}(\theta) = \langle \Phi^{[0]}|\,U^\dagger(\theta)\,H\, U(\theta)|\Phi^{[0]}\rangle, 
\end{equation}
or jointly as part of the combined GS+1QP cost functions described below for ground state and excited states. 

\subsubsection{Variance-based cost functions}

One effective approach is to minimize the energy variance \cite{Kardashin2020,zhang2022variational}, $\langle H^{2} \rangle - \langle H \rangle^{2}$. For a state $|\Psi\rangle = \sum_{n} c_{n} |\phi_{n}\rangle$ expanded in the eigenbasis of $H$ with eigenvalues $E_{n}$, the variance equals $\sum_{n} |c_{n}|^{2} (E_{n} - \bar{E})^{2}$, where $\bar{E} = \sum_{n} |c_{n}|^{2} E_{n}$ is the energy expectation value and the sum runs over energy eigenstates. The variance vanishes if and only if the state lies within a single energy eigenspace. For nondegenerate spectra, this requires an eigenstate; for degenerate subspaces, any superposition within the degenerate manifold yields zero variance. This quantity measures the extent to which $H$ has been block diagonalized and is related to residual off-diagonal elements in the context of continuous unitary transformations.

To decouple a quasidegenerate 1QP subspace from the rest of the spectrum, we generalize to multiple states. For a set of 1QP basis states $\{|\Phi^{[1]}_{i}\rangle\}_{i=1}^{N}$ transformed by unitary $U(\theta)$, the variance cost function is
\begin{eqnarray}\label{eq:C_var}
C_{\mathrm{var}}^{1\mathrm{QP}}(\theta)
  &=& \sum_{i=1}^{N} \langle \Phi^{[1]}_{i}|\,U^\dagger(\theta)\,H^{2}\,U(\theta)|\Phi^{[1]}_{i} \rangle \\ \nonumber
        &-& \sum_{i,j=1}^{N} \bigl|\langle \Phi^{[1]}_{i} |\,U^\dagger(\theta)\,H\,U(\theta)|\Phi^{[1]}_{j}\rangle\bigr|^{2}\, .
\end{eqnarray}
This cost function is unitarily invariant within the 1QP subspace: Any rotation of the basis $\{|\Phi^{[1]}_{i}\rangle\}$ by a unitary acting only within this subspace leaves $C^{\mathrm{1QP}}_{\mathrm{var}}$ unchanged. Physically, this means the cost function enforces block-diagonalization (decoupling the 1QP subspace from other sectors) without requiring full diagonalization within the 1QP block itself; the internal structure of the block remains flexible. In contrast, minimizing the variance for each state individually (as done in Ref.~\cite{zhang2022variational}) would not preserve this unitary invariance in the full subspace. The residual variance after optimization directly indicates the quality of decoupling achieved.

We can use variance-based cost functions in two ways. Either optimize $C_{\mathrm{var}}^{\mathrm{1QP}}$ alone for the 1QP sector (with a separate ground-state calculation using energy minimization) or include both sectors simultaneously:
\begin{eqnarray}\label{eq:cf_var_gs_1qp}
C^{\mathrm{GS, 1QP}}_{\mathrm{var}}(\theta) &=&  \langle \Phi^{[0]} |\,U^\dagger(\theta)\,H^{2}\,U(\theta)|{\Phi^{[0]}}\rangle \\ \nonumber
&-& |\langle \Phi^{[0]} |\,U^\dagger(\theta)\,H\,U(\theta)|{\Phi^{[0]}}\rangle|^{2} +  C^{\mathrm{1QP}}_{\mathrm{var}} (\theta),
\end{eqnarray}
where $|\Phi^{[0]}\rangle$ is the ground state of the unperturbed Hamiltonian $H_{0}$. We use equal weights (coefficient 1) for both the ground-state variance and the 1QP variance, treating both sectors symmetrically. Unlike the trace cost function (discussed below), which is unitarily invariant in the combined (GS+1QP) space, this sum has separate unitary invariances in each subspace. This has the advantage that minimizing variance in each sector separately actively drives orthogonality between the ground-state and 1QP subspace. Perfect optimization (zero variance) guarantees orthogonality; in practice, finite residual variance can reflect approximate orthogonality with the degree of mixing controlled by the residual.

\subsubsection{Trace-based cost functions}

An alternative approach minimizes the sum of weighted energies \cite{Nakanishi2019}:
\begin{eqnarray}\label{eq:C_trace}
C^{\mathrm{GS,1QP}}_{\mathrm{tr}}(\theta) &=& w_{0} \langle \Phi^{[0]}|\,U^\dagger(\theta)\,H\, U(\theta)|\Phi^{[0]}\rangle \\ \nonumber
&+& \sum_{i=1}^{N} w_{i} \langle \Phi^{[1]}_i|\,U^\dagger(\theta)\,H\, U(\theta)|\Phi^{[1]}_i\rangle.
\end{eqnarray}

We use uniform weights $w_{0} = w_{i} = 1$ for all $i$, making the cost function basis invariant. It is thus  the trace over the combined ground-state and 1QP subspace. This is natural for linked-cluster expansions where we require the full subspace information and treat all sites equivalently. We apply this trace cost function in two ways: $C_{\mathrm{tr}}^{\mathrm{1QP}}$ targets only the 1QP sector [$w_0 = 0$ in Eq.~\eqref{eq:C_trace} and separate ground-state optimization], while $C_{\mathrm{tr}}^{\mathrm{GS,1QP}}$ treats both ground-state and 1QP subspace with a single unitary.

The trace minimization is motivated by the fact that block diagonalization preserves the sum of eigenvalues within each block. However, when ground and excited states are included together, minimizing the trace does not guarantee that $U(\theta)|\Phi^{[0]}\rangle$ is the true lowest-energy state. A postprocessing diagonalization step (described in Sec.~\ref{sec:vqe:measurements}) is therefore required to identify the actual ground state and extract 1QP energies.

The trace cost function can also be evaluated using a purified approach where the whole subspace is treated as a wavefunction in a larger Hilbert space \cite{Benavides-Riveros2022,Guo2024}.\\

\subsubsection{Comparison and choice}

We distinguish between two different implementation strategies. With single-unitary approaches ($C_{\mathrm{tr}}^{\mathrm{GS,1QP}}$, $C_{\mathrm{var}}^{\mathrm{GS,1QP}}$), orthogonality is preserved throughout optimization by construction. However, $C_{\mathrm{tr}}^{\mathrm{GS,1QP}}$ requires postprocessing diagonalization to identify the true ground state, while $C_{\mathrm{var}}^{\mathrm{GS,1QP}}$ directly yields approximate eigenstates. With the separate-unitary approach ($C_{\mathrm{var}}^{\mathrm{1QP}}$), orthogonality must be verified and may require a generalized eigenvalue problem when non-orthogonality occurs.

We tested both strategies across all models studied in Sec.~\ref{sec:application}. For the pure TFIM cases, all cost functions achieve similar convergence and stability, successfully decoupling the low-energy block. The trace-based approach requires fewer measurements, as it only needs energy expectation values $\langle H \rangle$, while variance-based approaches additionally require $\langle H^2 \rangle$ measurements. Variance-based approaches provide residual variance as a diagnostic for block-diagonalization quality. However, for TFIM+LF where parity breaking makes optimization more challenging, trace-based cost functions show greater robustness to ground-state initialization, while variance-based approaches require near-zero initialization for larger clusters (see Sec.~\ref{sec:TFIM+LF}).

The variance-based approach $C_{\mathrm{var}}^{\mathrm{1QP}}$ has the additional flexibility of applying only to the excitation sector with separate ground-state calculation. However, with separate unitaries, orthogonality between ground-state and 1QP subspace is not enforced during optimization and must be verified. The other two cost functions ($C_{\mathrm{tr}}^{\mathrm{GS,1QP}}$ and $C_{\mathrm{var}}^{\mathrm{GS,1QP}}$) maintain orthogonality throughout optimization via their single-unitary structure. The residual value of $C_{\mathrm{var}}$ directly quantifies block-diagonalization quality, making it useful for convergence monitoring.

\subsection{State preparation}
\label{sec:vqe:state_prep}

The performance of VQE depends on the ansatz choice, which must balance expressibility and trainability. The HVA \cite{wecker_progress_2015} provides this balance by incorporating the problem structure directly into the circuit design. Motivated by quantum approximate optimization algorithm  \cite{Farhi2014QAOA} and adiabatic quantum computing \cite{farhi_quantum_2000}, HVA is formulated based on the structure of the Hamiltonian under investigation and has been successfully applied for ground-state calculations in quantum many-body systems \cite{wecker_progress_2015, wiersema_exploring_2020, wang_bang-bang_2022, AnselmeMartin2022, mele_avoiding_2022, Park2024}.

\subsubsection{Hamiltonian variational ansatz}

Our recent work \cite{Sumeet2024} investigated HVA performance for ground-state energy calculations. Here, we employ HVA to minimize the cost functions in Eqs.~\eqref{eq:C_var} and \eqref{eq:C_trace} that block diagonalize the Hamiltonian. For a Hamiltonian decomposed as $H=\sum_{\mu} H_{\mu}$, the HVA unitary takes the form
\begin{equation}
U_{\mathrm{HVA}}(\theta) = \prod_{l} \big(\prod_{\mu} e^{\mathrm{i}\theta_{l,\mu} H_{\mu}} \big),
\end{equation}
where $l$ indexes the layers. The HVA construction follows the physical intuition of adiabatic evolution: Each layer implements a discrete evolution step alternating between the unperturbed Hamiltonian $H_0$ and the perturbation $V$, with variational parameters that can transform the unperturbed eigenstates into eigenstates of the full Hamiltonian $H = H_0 + xV$. Following the bang-bang digitized adiabatic approach \cite{wang_bang-bang_2022}, we group commuting terms together and separate non-commuting terms to create blocks that are individually classically solvable, thereby minimizing Trotter error.

\begin{figure}
    \centering
    \includegraphics[width=1\linewidth]{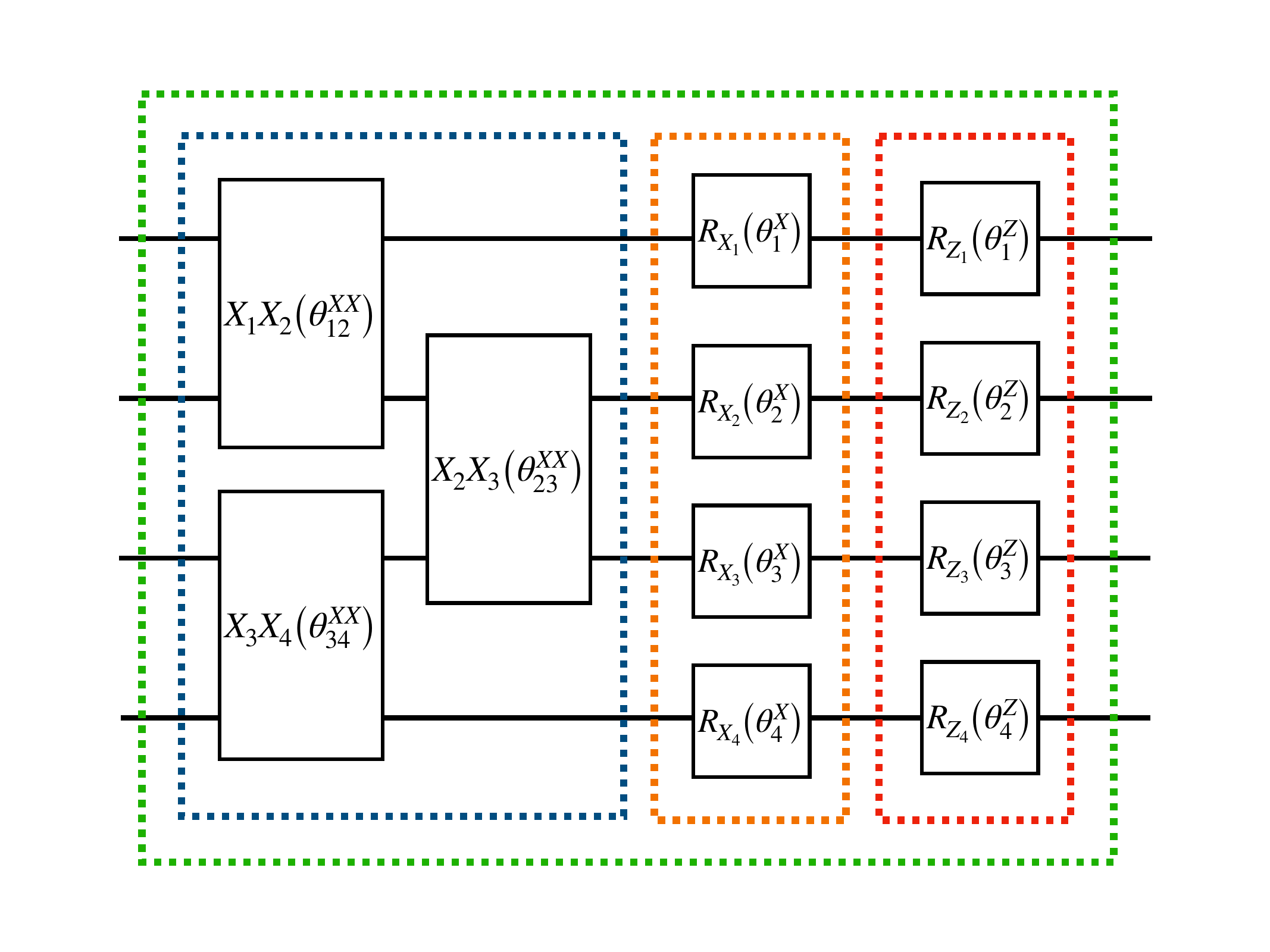}
    \caption{Quantum circuit representation of one layer (green box) of the HVA for the one-dimensional TFIM+LF model. The blue, orange, and red blocks implement $e^{\mathrm{i}\sum_{\langle\nu,\mu\rangle}\theta^{XX}_{\nu,\mu}X_{\nu} X_{\mu}}$, $e^{\mathrm{i}\sum_{\nu}\theta^{X}_{\nu} X_{\nu}}$, and $e^{\mathrm{i}\sum_{\nu}\theta^{Z}_{\nu} Z_{\nu}}$, respectively, showing the $XX-X-Z$ ordering within each layer.} 
    \label{fig:HVA_circuit}
\end{figure}

For the TFIM+LF [Eq.~\eqref{eq:ham_overview}] on any cluster graph $C$, the HVA unitary is:
\begin{equation}\label{eqn:hva}
U_{\mathrm{HVA},\, C}(\theta) = \prod_{l=1}^{N_{l}} e^{\mathrm{i}\sum_{\nu}\theta^{Z}_{\nu,l} Z_{\nu}} \cdot e^{\mathrm{i}\sum_{\nu}\theta^{X}_{\nu,l} X_{\nu}} \cdot e^{\mathrm{i}\sum_{\langle\nu,\mu\rangle}\theta^{XX}_{\nu,\mu,l}X_{\nu} X_{\mu}}.
\end{equation}
The $XX-X-Z$ ordering exemplifies this strategy: Since $[X_{\nu} X_{\mu}, X_{\rho}] = 0$, the $XX$-coupling and $X$-field blocks commute and are placed adjacently, forming  $H_{XX-X} = -\sum_{\langle\nu,\mu\rangle} X_{\nu} X_{\mu} - h_{X} \sum_{\nu} X_{\nu}$, which is classically solvable (eigenstates are product states in the $X$ basis). The $Z$-field block $H_{Z} = -h_{Z} \sum_{\nu} Z_{\nu}$ is also diagonal. This creates two classically solvable parts per layer. On a cluster of $N$ spins, each HVA layer requires $N$ qubits, $2N$ single-qubit rotations ($N$ for the $Z$ block and $N$ for the $X$ block), and one two-qubit gate per nearest-neighbor bond from the $XX$ block ($N-1$ for an open one-dimensional chain). This layer structure is illustrated in Fig.~\ref{fig:HVA_circuit} for the one-dimensional case. The general principle of grouping commuting terms to form classically solvable blocks applies to arbitrary cluster geometries.

%\paragraph{Symmetries and parameter reduction:}
We emphasize that the presence of symmetries allows for parameter reduction. Symmetries of the Hamiltonian translate into symmetries of the optimal unitary, constraining the variational parameters. If $[H,\mathcal{S}] = 0$ for symmetry operation $\mathcal{S}$, then any block-diagonalizing unitary $U$ can be chosen to commute with $\mathcal{S}$, imposing parameter constraints. For the one-dimensional chain, reflection symmetry about the center implies $\theta^{q}_{i} = \theta^{q}_{N+1-i}$ (where $q$ labels the parameter type), reducing parameters by roughly half. For NLCE, most clusters have only discrete symmetries (reflection, rotation); periodic chains additionally have translational invariance allowing further reduction.

\subsubsection{Initialization strategy}

Good initialization is crucial for successful optimization. We employ a two-stage protocol. All qubits are initialized in the computational basis state $|0\rangle$, denoted as $|\mathrm{ref}\rangle$. For the high-field polarized phase of the TFIM, this coincides with the unperturbed ground state $|\Phi^{[0]}\rangle$ of $H_0$. The initial parameters $\theta_{\mathrm{initial}}$ determine the starting configuration via $U_\mathrm{HVA}(\theta_{\mathrm{initial}})|\mathrm{ref}\rangle$.\\

\textit{Stage 1} We first optimize $U_\mathrm{HVA}(\theta)$ for the ground state alone by minimizing energy expectation value (not variance), ensuring convergence to the lowest-energy state via the Rayleigh-Ritz variational principle. Initializing all parameters to small values close to zero ensures reliable convergence.

\textit{Stage 2} We use the optimized ground-state parameters $\theta_{\mathrm{GS}}$ as initialization for optimizing the full subspace cost functions:
\begin{equation}
U_\mathrm{HVA}(\theta) X_{i}|\mathrm{ref}\rangle = U_\mathrm{HVA} (\theta) X_{i} U_\mathrm{HVA}(\theta_{\mathrm{GS}})^\dagger U_\mathrm{HVA}(\theta_{\mathrm{GS}}) |\mathrm{ref}\rangle,
\end{equation}
where $X_{i}|\mathrm{ref}\rangle$ form the 1QP basis of $H_{0}$. If the ground-state unitary provides a good approximation to the true ground state, then $U_\mathrm{HVA}(\theta_{\mathrm{GS}}) X_{i} U_\mathrm{HVA}(\theta_{\mathrm{GS}})^\dagger$ acts as an approximate excitation operator. For the one-dimensional TFIM, this initialization already provides nearly exact 1QP states, requiring minimal further optimization. For two-dimensional TFIM and models with longitudinal field, the initialization provides a good starting point that improves convergence compared to random parametrization, though substantial optimization is still required to achieve accurate decoupling.\\

The ground-state solution provides a physics-informed initialization strategy that leverages the already-optimized ground-state parameters. This approach reflects warm-start principles, beginning optimization with lower initial cost function values compared to random initialization, and proves effective in most cases. However, as we show in  Secs.~\ref{sec:2D} and \ref{sec:TFIM+LF}, it can lead to local minima for variance-based cost functions when optimizing larger clusters, while trace-based approaches remain robust. In cases where convergence issues arise, initializing parameters near zero provides more stable convergence despite starting with higher initial cost function values.
\subsection{Quantum circuit implementation and measurement requirements}
\label{sec:vqe:measurements}

Once VQE optimization is complete and we have obtained the optimized parameters $\theta_{\mathrm{opt}}$, we must perform additional quantum measurements to extract the information needed for constructing the PCAT unitary $U_{\mathrm{PCAT}}$ and the effective 1QP Hamiltonian $H_{\mathrm{eff}}^{[1]}$. As discussed in Sec.~\ref{sec:PCAT}, PCAT requires constructing modified states that eliminate projections onto lower-energy subspaces of $H_{0}$. Since VQE provides a unitary transformation $U_{\mathrm{VQE}}$ rather than explicit state vectors, we must extract the necessary information through quantum measurements.

We organize this procedure into two parts: Sec.~\ref{sec:measurements} describes the quantum measurements required to obtain three essential $(N+1) \times (N+1)$ matrices ($\mathcal{O}$, $S$, $\tilde{H}$), and Sec.~\ref{sec:hardware} describes the classical postprocessing steps to construct $U_{\mathrm{PCAT}}$ and extract $H_{\mathrm{eff}}^{[1]}$.

\subsubsection{Quantum measurements}\label{sec:measurements}

After VQE optimization yields optimal parameters $\theta_{\mathrm{opt}}$, we have a set of $N+1$ VQE-prepared states that approximately span the ground-state and 1QP subspace.  We label the VQE-prepared ground state as $|\chi^{[0]}\rangle$ and the VQE-prepared 1QP states as $|\chi^{[1]}_{i}\rangle$ for $i=1,\ldots,N$. We use $\chi$ to denote VQE-prepared states, reserving $\Psi$ for exact eigenstates. Here, $U^{[0]}_{\mathrm{VQE}} = U_{\mathrm{HVA}}(\theta_{\mathrm{GS}})$ and $U^{[1]}_{\mathrm{VQE}} = U_{\mathrm{HVA}}(\theta_{\mathrm{1QP}})$, where $\theta_{\mathrm{GS}}$ and $\theta_{\mathrm{1QP}}$ are the optimized parameters for ground state and excited states. $U^{[0,1]}_{\mathrm{VQE}} = U_{\mathrm{HVA}}(\theta_{\mathrm{GS,1QP}})$ is used for the common unitary obtained after the optimization of the cost functions in Eqs.(~\ref{eq:cf_var_gs_1qp}) and (~\ref{eq:C_trace}). \\

How these states are prepared depends on the cost function used:
\begin{itemize}
\item \textit{For} $C^{\mathrm{GS,1QP}}_{\mathrm{tr}}$ \textit{or} $C_{\mathrm{var}}^{\mathrm{GS,1QP}}$ \textit{(single unitary)}: All states use the same optimized unitary: $|\chi^{[0]}\rangle = U^{[0,1]}_{\mathrm{VQE}}|\Phi^{[0]}\rangle$, $|\chi^{[1]}_{i}\rangle = U^{[0,1]}_{\mathrm{VQE}}|\Phi^{[1]}_{i}\rangle$, where $|\Phi^{[0]}\rangle$ is the unperturbed ground state (which equals $|\mathrm{ref}\rangle$ for our models) and $|\Phi^{[1]}_{i}\rangle = X_{i}|\mathrm{ref}\rangle$ are the unperturbed 1QP states.

\item \textit{For} $C_{\mathrm{var}}^{\mathrm{1QP}}$ \textit{with separate ground-state optimization}: We use different unitaries: $|\chi^{[0]}\rangle = U^{[0]}_{\mathrm{VQE}}|\Phi^{[0]}\rangle$, $|\chi^{[1]}_{i}\rangle = U^{[1]}_{\mathrm{VQE}}|\Phi^{[1]}_{i}\rangle$.
\end{itemize}

The three essential $(N+1) \times (N+1)$ matrices we must measure are:\\

\textit{Matrix $\mathcal{O}$ (overlap with unperturbed states)} This $(N+1) \times (N+1)$ matrix contains overlaps between VQE-prepared states (rows) and unperturbed eigenstates $|\Phi^{[0]}_j\rangle$ (columns, $j \in \{0,1,\ldots,N\}$). The ground-state row is $\mathcal{O}_{0j} = \langle\chi^{[0]}|\Phi^{[0]}_j\rangle$, and the 1QP rows are $\mathcal{O}_{ij} = \langle\chi^{[1]}_{i}|\Phi^{[0]}_j\rangle$ for $i=1,\ldots,N$.\\

\textbf{Matrix $S$ (overlap between VQE-prepared states)}
This matrix contains overlaps between the VQE-prepared states themselves. Elements are $S_{00} = \langle\chi^{[0]}|\chi^{[0]}\rangle$, $S_{0i} = \langle\chi^{[0]}|\chi^{[1]}_{i}\rangle$, and $S_{ij} = \langle\chi^{[1]}_{i}|\chi^{[1]}_{j}\rangle$.

For single-unitary cost functions, $S = \mathds{1}_{N+1}$ exactly (diagonal by construction). For separate-unitary approaches, off-diagonal elements $S_{0i} = \langle\chi^{[0]}|\chi^{[1]}_{i}\rangle$ will be nonzero (unless there is an additional symmetry as in the pure TFIM), reflecting imperfect orthogonality.\\

\textbf{Matrix $\tilde{H}$ (Hamiltonian expectation values)}
This matrix contains Hamiltonian matrix elements in the basis of VQE-prepared states. Elements are $\tilde{H}_{00} = \langle\chi^{[0]}|H|\chi^{[0]}\rangle$, $\tilde{H}_{0i} = \langle\chi^{[0]}|H|\chi^{[1]}_{i}\rangle$, and $\tilde{H}_{ij} = \langle\chi^{[1]}_{i}|H|\chi^{[1]}_{j}\rangle$.\\

%\textbf{Measurement protocols.}
The overlap matrix $\mathcal{O}$ and (when needed) the overlap matrix $S$ are measured using the Hadamard test \cite{Nielsen_Chuang_2010}. The Hamiltonian matrix $\tilde{H}$ can be measured more directly: Diagonal elements $\tilde{H}_{00}$ and $\tilde{H}_{ii}$ are expectation values obtained via standard Pauli measurements, while off-diagonal elements can be measured as expectation values on superposition states when both states are prepared using the same unitary. Specifically, by preparing superposition states such as $|\chi_{\pm}\rangle = (|\chi^{[0]}\rangle \pm |\chi^{[1]}_{i}\rangle)/\sqrt{2}$ or $(|\chi^{[1]}_{i}\rangle \pm |\chi^{[1]}_{j}\rangle)/\sqrt{2}$ and measuring $\langle H \rangle$, one obtains the real part of the off-diagonal elements; the imaginary part is obtained similarly using complex superpositions.

For separate-unitary approaches, overlaps and Hamiltonian matrix elements that couple different unitaries require the Hadamard test with composite unitaries. For example, $S_{0i} = \langle\chi^{[0]}|\chi^{[1]}_{i}\rangle = \langle\Phi^{[0]}|U_{\mathrm{VQE}}^{[0]\dagger} U^{[1]}_{\mathrm{VQE}}|\Phi^{[1]}_{i}\rangle$ requires Hadamard test. Similarly, Hamiltonian expectation values coupling different unitaries, such as $\tilde{H}_{0i} = \langle\chi^{[0]}|H|\chi^{[1]}_{i}\rangle = \langle\Phi^{[0]}|U_{\mathrm{VQE}}^{[0]\dagger} H U^{[1]}_{\mathrm{VQE}}|\Phi^{[1]}_{i}\rangle$, also require Hadamard tests with composite unitaries. Within the 1QP block using the same unitary, $\tilde{H}_{ij} = \langle\chi^{[1]}_{i}|H|\chi^{[1]}_{j}\rangle$ can use superposition state measurements.

The total number of measurements scales as $O(N^{2})$ for each matrix. However, these measurements are performed only once after VQE optimization is complete, so they do not represent the computational bottleneck; the iterative VQE optimization itself is far more resource demanding.

\subsubsection{Classical postprocessing}\label{sec:hardware}

Having obtained the three matrices $\mathcal{O}$, $S$, and $\tilde{H}$ from quantum measurements, we now proceed entirely classically to construct $U_{\mathrm{PCAT}}^{[1]}$ and extract $H_{\mathrm{eff}}^{[1]}$. This construction proceeds in four steps.\\

\textit{Step 1: Identify the true ground state and handle non-orthogonality.}

When using $C^{\mathrm{GS,1QP}}_{\mathrm{tr}}$, the optimization does not guarantee that the nominal ground state $|\chi^{[0]}\rangle$ is the true lowest-energy eigenstate. When using separate unitaries ($C_{\mathrm{var}}^{\mathrm{1QP}}$), there may be non-zero overlaps $S_{0i} \neq 0$ between ground and 1QP states. In both cases, we must solve an eigenvalue problem to find orthogonal eigenstates.

For single-unitary cases where $S = \mathds{1}_{N+1}$, we diagonalize $\tilde{H}$:
\begin{equation}
\tilde{H} = W D W^\dagger,
\end{equation}
where $W$ is the unitary matrix of eigenvectors and $D$ is the diagonal matrix of eigenvalues.

For separate-unitary cases where $S \neq \mathds{1}_{N+1}$, we solve the generalized eigenvalue problem:
\begin{equation}
\tilde{H} v = \lambda S v,
\end{equation}
which can be transformed to standard form and yields eigenvectors that form the columns of matrix $W$. The lowest (most negative) eigenvalue identifies the true ground state.

This transformation redefines our basis to the true orthonormal eigenstates. The true ground state is:
\begin{equation}
|\chi^{[0]}_{\mathrm{true}}\rangle = W_{00}|\chi^{[0]}\rangle + \sum_{i=1}^{N} W_{0i}|\chi^{[1]}_{i}\rangle,
\end{equation}
and the true 1QP eigenstates are
\begin{equation}
|\chi^{[1]}_{i,\mathrm{true}}\rangle = W_{i0}|\chi^{[0]}\rangle + \sum_{j=1}^{N} W_{ij}|\chi^{[1]}_{j}\rangle \quad \text{for } i=1,\ldots,N.
\end{equation}
The matrices in this new basis are
\begin{eqnarray}
\mathcal{O}_{\mathrm{true}} &=& W^\dagger \mathcal{O},\\ \nonumber
\quad \tilde{H}_{\mathrm{true}} &=& W^\dagger \tilde{H} W = D,\\ \nonumber
\quad S_{\mathrm{true}} &=& W^\dagger S W = \mathds{1}_{N+1},
\end{eqnarray}
where $S_{\mathrm{true}} = \mathds{1}_{N+1}$ confirms that the transformed states are orthonormal. For the generalized eigenvalue problem with non-orthogonal initial states ($S \neq \mathds{1}_{N+1}$), the matrix $W$ obtained from the eigenvalue solver is not unitary. The transformation can be equivalently understood as a two-step process: First preorthogonalize the basis using the symmetric orthogonalization $(S^{\dagger}S)^{-1/2}$, and then apply a unitary transformation to diagonalize the Hamiltonian in the orthogonal basis. The generalized eigenvalue solver performs both steps simultaneously, directly producing orthonormal eigenvectors with respect to the metric $S$.

For single-unitary variance cost functions ($C_{\mathrm{var}}^{\mathrm{GS,1QP}}$) with successful optimization, $S = \mathds{1}_{N+1}$ already, but diagonalizing $\tilde{H}$ can still improve state quality by finding exact eigenstates within the VQE-prepared subspace. 

We proceed to Step 2 using the transformed matrices. For notational simplicity, we drop the $_{\mathrm{true}}$ subscript and continue to denote the VQE-prepared states as $|\chi^{[0]}\rangle$ and $|\chi^{[1]}_{i}\rangle$, understanding these now refer to the orthonormalized eigenstates. The matrices $\mathcal{O}$, $\tilde{H}$, and $S$ refer to these corrected, orthonormalized versions.\\

\textit{Step 2: Construct modified overlap matrix using PCAT recursion.}

According to the PCAT prescription (Sec.~\ref{sec:PCAT}), modified states eliminate projections onto lower-energy subspaces of $H_{0}$. For the 1QP sector, the recursion from Eq.~\eqref{eq:pcat-recursion-1qp} is

\begin{equation}
|\widetilde{\chi}^{[1]}_{i}\rangle = |\chi^{[1]}_{i}\rangle - \frac{\langle\Phi^{[0]}|\chi^{[1]}_{i}\rangle}{\langle\Phi^{[0]}|\chi^{[0]}\rangle}|\chi^{[0]}\rangle
\end{equation}
for $i = 1, \ldots, N$. Taking the overlap with an unperturbed state $|\Phi^{j}\rangle$ yields
\begin{eqnarray}
\langle \Phi^{[0]}_j|\widetilde{\chi}^{[1]}_{i}\rangle &=& \langle\Phi^{[0]}_j|\chi^{[1]}_{i}\rangle - \frac{\langle\Phi^{[0]}|\chi^{[1]}_{i}\rangle}{\langle\Phi^{[0]}|\chi^{[0]}\rangle}\langle \Phi^{[0]}_{j}|\chi^{[0]}\rangle \\ \nonumber
&=& \mathcal{O}_{ij} - \frac{\mathcal{O}_{i0}}{\mathcal{O}_{00}} \mathcal{O}_{0j}.
\end{eqnarray}

The modified overlap matrix $\widetilde{\mathcal{O}}$ for the 1QP sector is an $N \times (N+1)$ matrix:
\begin{equation}
\widetilde{\mathcal{O}}_{ij} = \mathcal{O}_{ij} - \frac{\mathcal{O}_{i0}}{\mathcal{O}_{00}} \mathcal{O}_{0j}
\end{equation}
for rows $i=1,\ldots,N$ (1QP states) and columns $j=0,1,\ldots,N$ (all unperturbed states). Restricting to the 1QP subspace of $H_{0}$ (columns $j=1,\ldots,N$), we obtain the $N \times N$ modified overlap matrix:
\begin{equation}
\widetilde{X}^{[1]}_{ik} = \mathcal{O}_{i,k} - \frac{\mathcal{O}_{i0}}{\mathcal{O}_{00}} \mathcal{O}_{0,k}
\end{equation}
for $i,k = 1, \ldots, N$. This ensures that $P^{[0]}|\widetilde{\chi}^{[1]}_{i}\rangle = 0$ for all 1QP states.\\

\textit{Step 3: Construct the PCAT correction unitary.}

The PCAT correction within the 1QP subspace is the $N \times N$ unitary matrix:
\begin{equation}
V^{[1]} = \widetilde{X}^{[1]\dagger} (\widetilde{X}^{[1]} \widetilde{X}^{[1]\dagger})^{-1/2}.
\end{equation}
This corresponds to the transformation described in Sec.~\ref{sec:PCAT}, where the full PCAT unitary for the 1QP block is $U_{\mathrm{PCAT}}^{[1]} = U^{[1]} V^{[1]}$, with $U^{[1]}$ being the matrix whose columns are the VQE-approximated 1QP eigenstates. Here, $V^{[1]}$ represents the additional correction needed beyond the VQE unitary to enforce cluster additivity.\\

\textit{Step 4: Extract the effective 1QP Hamiltonian}\\

The effective 1QP Hamiltonian is obtained by restricting the Hamiltonian matrix (from Step 1) to the 1QP sector and applying the PCAT correction. We extract the $N \times N$ subblock corresponding to the 1QP sector (rows and columns with indices $1,\ldots,N$) from the $(N+1) \times (N+1)$ transformed Hamiltonian. This restriction removes the ground state (index 0) and isolates the 1QP subspace:
\begin{equation}
\tilde{H}^{[1]}_{ij} = \tilde{H}_{i,j} \quad \text{for } i,j \in \{1,\ldots,N\}.
\end{equation}

The effective 1QP Hamiltonian is then
\begin{equation}
H_{\mathrm{eff}}^{[1]} = V^{[1]\dagger} \tilde{H}^{[1]} V^{[1]} - E^{[0]}.
\end{equation}

This polynomial-time classical computation transforms the VQE output, encapsulated in the three matrices $\mathcal{O}$, $S$, and $\tilde{H}$, into the cluster-additive effective Hamiltonian required for NLCE. The resulting $H_{\mathrm{eff}}^{[1]}$ is an $N \times N$ matrix in the 1QP basis, which after subtracting the ground-state energy $E^{[0]}$, gives the effective 1QP Hamiltonian for the cluster as defined in Sec.~\ref{sec:LCE}:
\begin{equation}\label{eq:final_h_eff}
H^{[1]}_{\mathrm{eff}, C} = (U_{\mathrm{PCAT}, C}^\dagger H U_{\mathrm{PCAT}, C})^{[11]} - E^{[0]}_{C}
\end{equation}
This quantity is then used in the NLCE embedding procedure [Eq.~\eqref{eq:H_bar}] to obtain the 1QP dispersion in the thermodynamic limit.

\section{Applications}\label{sec:application}

We now apply the NLCE+VQE approach of the previous sections to compute 1QP excitation energies in the TFIM. We investigate the one-dimensional chain at the critical point, the two-dimensional square lattice at its quantum critical point, and additionally the one-dimensional TFIM with longitudinal field (TFIM+LF) in the disordered phase. The longitudinal field fundamentally changes the model by breaking parity symmetry, coupling ground state to excitations, and making the one-dimensional model nonintegrable. 

To benchmark NLCE+VQE, we compare its performance with NLCE$+$ED, where ED is used to compute properties on finite clusters. Furthermore, we investigate whether the variance- and trace-based cost functions yield similar results or not. In all three cases, we use the conjugate gradient method for the classical optimization in the VQE process.

\subsection{Transverse-field Ising model}

The 1D TFIM is one of the paradigmatic models for studying
quantum phase transitions driven by quantum fluctuations. 
The Hamiltonian of the TFIM is given by
\begin{equation}\label{ham_tfim}
    H = - h \sum_\nu Z_\nu - J \sum_{<\nu^\prime,\nu>} X_{\nu^\prime} X_{\nu^{\phantom\prime}},
\end{equation}
where $Z_\nu$ and $X_\nu$ denote the Pauli matrices acting on site $\nu$, 
$J$ represents the nearest-neighbor Ising exchange coupling, and $h$ the strength of the transverse field. Throughout this work, we set $h = 1$ as our energy unit, such that all energies are measured in units of the transverse-field strength, and we consider ferromagnetic $J>0$. For $h = 0$, the ground state is doubly degenerate, corresponding to ferromagnetic order along the $x$ axis and spontaneous breaking of the global $\mathbb{Z}_2$ symmetry generated by $\prod_\nu X_\nu$. In contrast, for $J=0$, the ground state is a product state of spins polarized along the $z$ direction, forming a paramagnetic phase that preserves $\mathbb{Z}_2$ symmetry. The competition between these two noncommuting terms leads to a quantum critical point at $J_c$ separating the ordered and disordered phases.

The TFIM in one dimension is exactly solvable via Jordan-Wigner transformation \cite{pfeuty1970}, making it an ideal starting benchmark for validating our NLCE+VQE framework.
In higher dimensions, the TFIM becomes nonintegrable but retains the same qualitative features.
We perform expansions from the high-field polarized phase where the ground state is gapped and 1QP excitations correspond to localized spin flips, and study the quantum critical point where the gap closes.

\subsubsection{One-dimensional chain}\label{sec:1D}

The one-dimensional TFIM undergoes a second-order phase transition at $J_c=1$. The system is disordered when $J<1$ and ordered (Ising ferromagnetic) when $J>1$. We perform an expansion from the disordered phase as the unperturbed limit and present NLCE+VQE results at the critical point.

\begin{figure}
    \centering
    \includegraphics[width=0.95\linewidth]{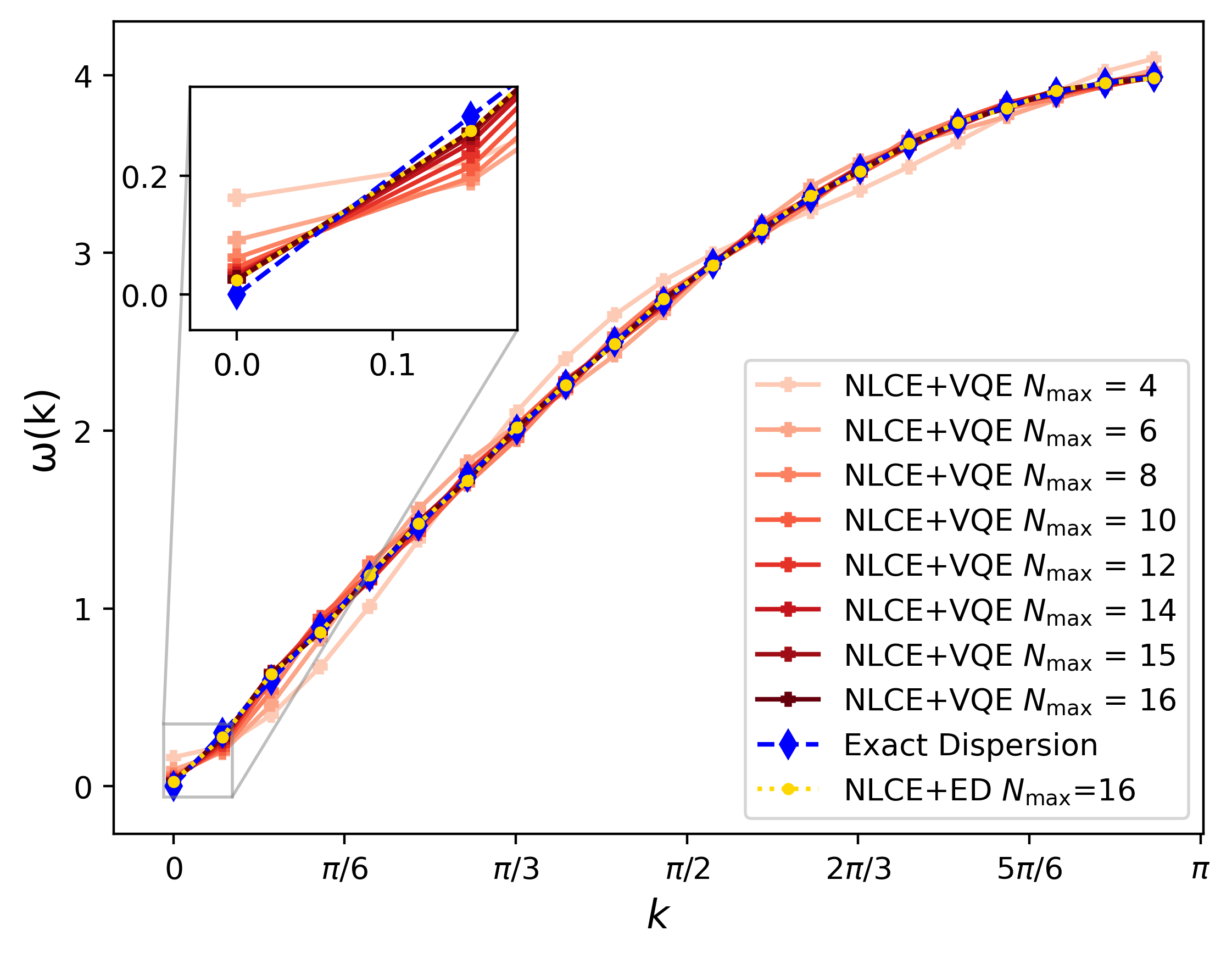}
    \caption{1QP dispersion $\omega(k)$ for the 1D TFIM at the critical point $J_c=1$ in the thermodynamic limit. Gradient of red lines with plus signs shows NLCE+VQE results for different maximum cluster sizes, yellow dotted line with circles shows NLCE+ED for $N_{\mathrm{max}}=16$, and blue dotted line with diamonds shows the exact solution~\cite{pfeuty1970}. The inset zooms into the region around $k=0$ where the gap closes. Results use $\lceil N/2 \rceil$ HVA layers for finite clusters in NLCE+VQE.}
    \label{fig:1d_results}
\end{figure}

The exact 1QP dispersion in the thermodynamic limit is given by

\begin{equation}
    \omega_{\mathrm{exact}}(k)= 2 \sqrt{1+ J^2 -2J \cos{(k)}},
\end{equation}

where $k$ labels the momentum quantum number of the 1QP state, $k \in [-\pi,\pi]$. At the critical point $J_c = 1$, the excitation gap closes at $k=0$, resulting in a linear dispersion $\omega(k) \propto |k|$ characteristic of the $(1+1)$-dimensional Ising universality class. This vanishing gap at criticality provides a stringent test of NLCE convergence.

For the one-dimensional chain, rectangular graphs have dimensions $L_m\times 1 \leq N_{\mathrm{max}}$. The embedding factors in NLCE are equal to $1$. Hence, the expansion reduces to the contributions from only two clusters corresponding to $N$ and $N-1$ and the terms corresponding to smaller clusters cancel out. The dispersion $\omega(k)$ in the momentum states $k$ for one-dimensional chain can thus be written as 

\begin{eqnarray}\label{eq:1d_nlce}
    \omega(k) &=& \sum_{\nu,\mu}^N H^{[1]}_{\mathrm{eff},N,\nu,\mu} \, e^{{\rm i} \, k \,(\nu-\mu)} \\ \nonumber
    &-& \sum_{\nu',\mu'}^{N-1} H^{[1]}_{\mathrm{eff},N-1,\nu'\mu'} \, e^{{\rm i} \, k \, (\nu'-\mu')},
\end{eqnarray}

where $H^{[1]}_{\mathrm{eff},N,\nu\mu}$ are the matrix elements of $H_{\mathrm{eff}}$ obtained with VQE for a hopping from site $\nu$ to site $\mu$ for a cluster $N$ given by $\langle \nu |H^{[1]}_{\mathrm{eff},N}|\mu \rangle$. The matrix $H^{[1]}_{\mathrm{eff},N}$ for a cluster of size $N$ is given in Eq.~\ref{eq:final_h_eff}.

\begin{figure}
    \centering
    \includegraphics[width=1\linewidth]{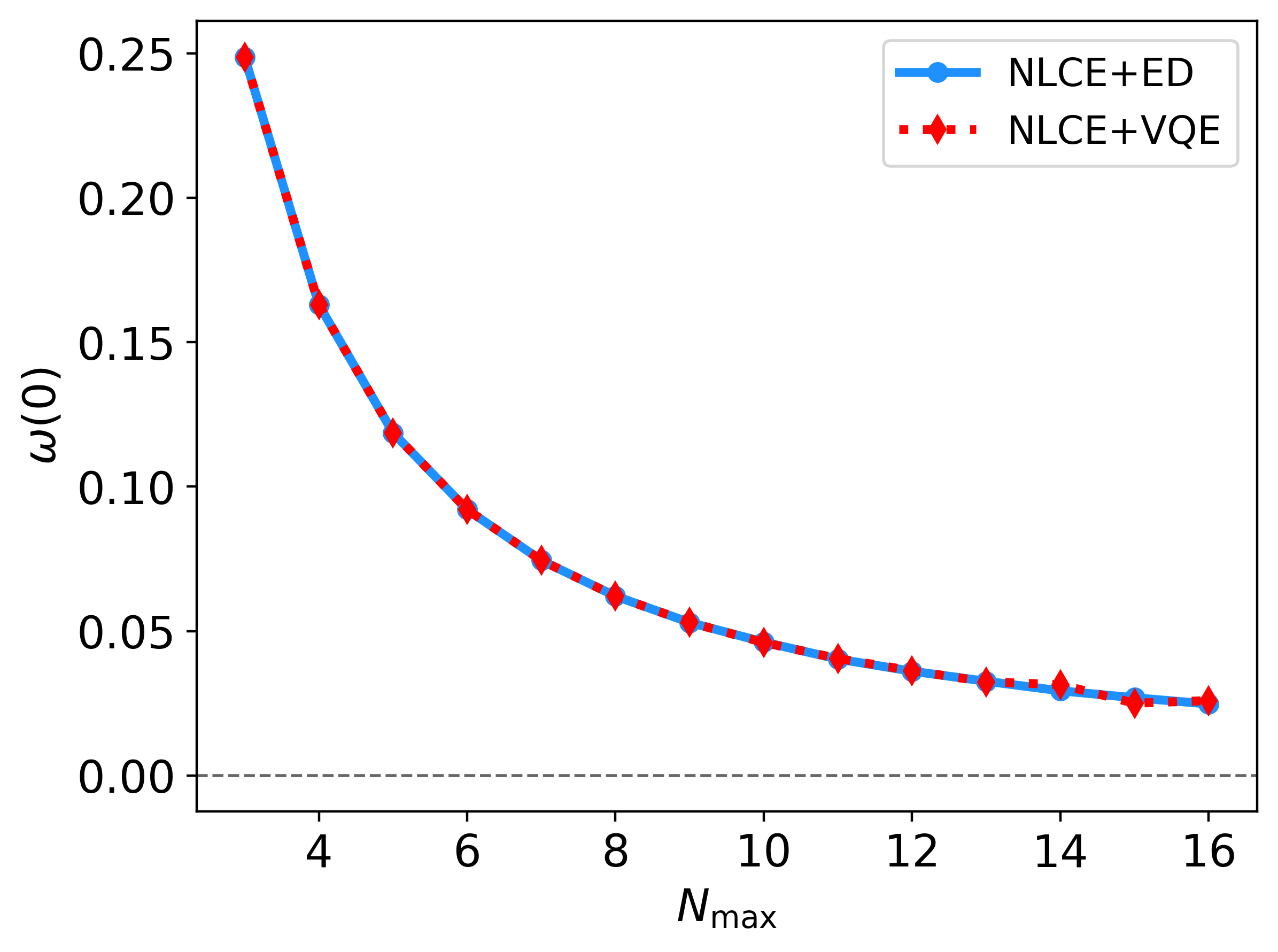}
    \caption{Convergence of NLCE with maximum cluster size for 1D TFIM at the critical point $J_c=1$ in the thermodynamic limit. The plot shows the dispersion $ \omega (0)$ at momentum $k=0$ obtained using NLCE+VQE (red dotted line with diamonds) and NLCE+ED (blue solid line with circles). Both exhibit identical algebraic convergence, demonstrating that $\lceil N/2 \rceil$ HVA layers suffice for the critical point.}
    \label{fig:1d_convergence}
\end{figure}

Remarkably, for the one-dimensional TFIM, the VQE unitary optimized solely to minimize ground-state energy automatically decouples all excitation sectors simultaneously. This is a special property of this model that warrants explanation. After the Jordan-Wigner transformation, the 1D TFIM maps to a free-fermion model described by a quadratic Hamiltonian. In such systems, a single Bogoliubov transformation diagonalizes the entire Hamiltonian across all quasiparticle sectors. The VQE unitary $U_{\mathrm{VQE}}$ effectively learns this Bogoliubov transformation when minimizing the ground-state energy. Importantly, while any block-diagonalization method must decouple the ground state from excitations, for a generic non-integrable system it would generally not decouple all higher quasiparticle sectors from each other. That the ground-state-optimized VQE unitary achieves complete block diagonalization is thus a remarkable feature of the free-fermion structure and integrability of this model, not a general expectation. This could explain why $\lceil N/2 \rceil$ HVA layers suffice for convergence: The target transformation may be efficiently expressible within the HVA ansatz structure.
\begin{figure*}[t]
    \centering
    \includegraphics[width=0.7\linewidth]{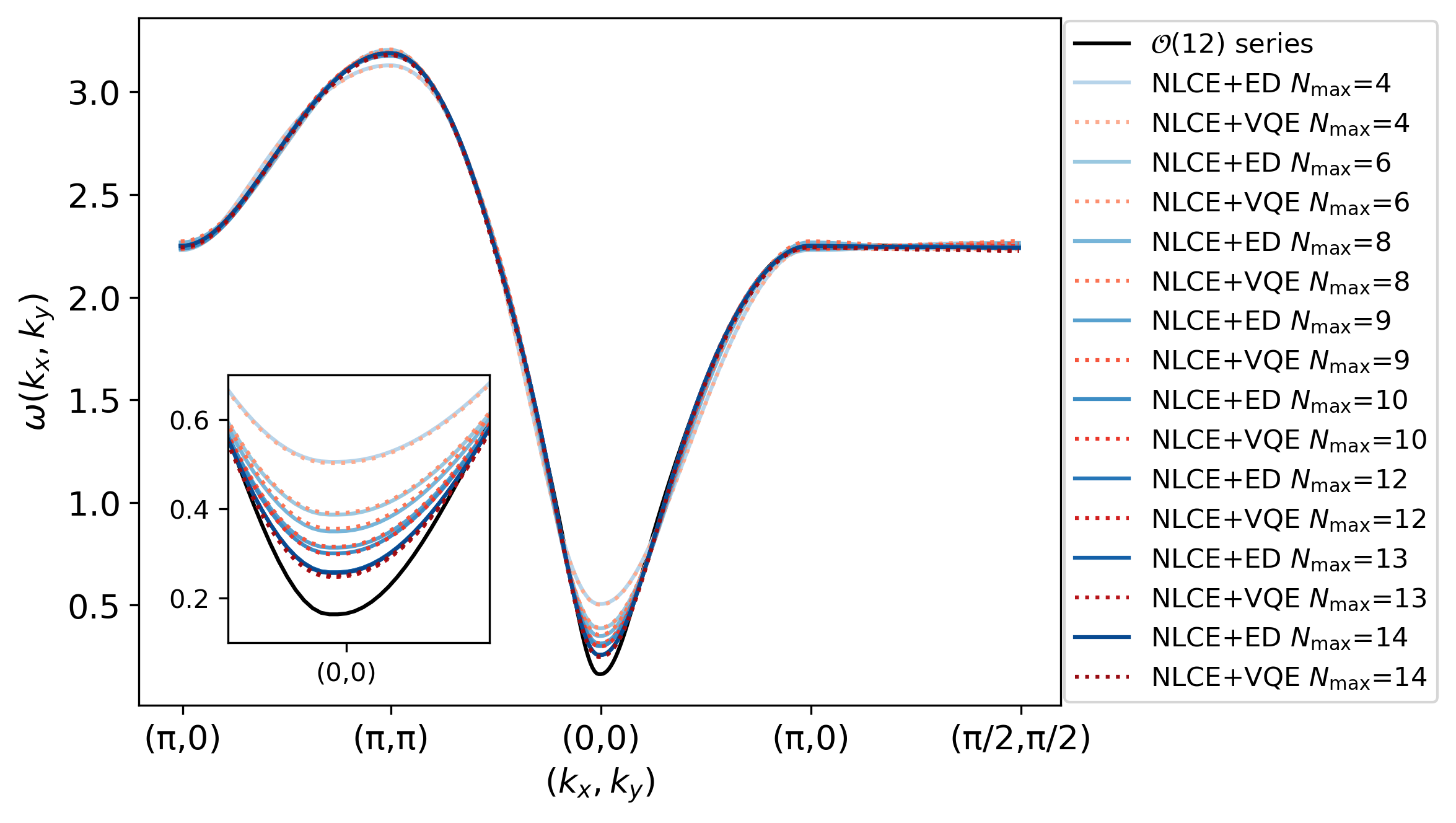}
    \caption{1QP dispersion $\omega(k_x, k_y)$ for the 2D TFIM on a square lattice at the quantum critical point $J_c=0.328$ \cite{He1990, Hesselmann2016}. Gradient of solid blue lines shows NLCE+ED results and gradient of dashed red lines shows NLCE+VQE results for different maximum cluster sizes $N_{\mathrm{max}}$. The inset shows an enlarged view of the critical region near $(k_x,k_y)=(0,0)$ where the gap closes. These results use $\lceil N/2 \rceil$ HVA layers for finite clusters in NLCE+VQE. The solid black line refers to the bare series expansion of order 12 about the high-field limit.}
    \label{fig:2d_critical_dispersion}
\end{figure*}

In practice, we compute the ground-state parameters with NLCE+VQE using $\lceil N/2 \rceil$ layers of HVA and then construct the 1QP subspace with this unitary. We performed statevector simulations for 1D TFIM. Figure~\ref{fig:1d_results} shows the resulting dispersion $\omega(k)$ for cluster sizes up to $N_{\mathrm{max}} = 16$, where $N_{\mathrm{max}}$ denotes the largest cluster included in the NLCE expansion. As shown in the figure, one can observe that as the cluster size increases, the NLCE+VQE results converge rapidly toward both the NLCE+ED (for $N_{\mathrm{max}} = 16$) and the exact dispersion, confirming the validity of the convergence of NLCE with implementation of variational approach for finite clusters. The inset of Fig.~\ref{fig:1d_results} highlights the region near $k = 0$, the critical momentum where the gap vanishes at $J_c = 1$. Despite the vanishing gap making NLCE convergence more demanding, the progressive approach of NLCE+VQE data toward the exact curve with increasing cluster size demonstrates that the method correctly captures the critical behavior and the linear dispersion $\omega(k) \propto |k|$ near the gapless point. 

Quantitative convergence is illustrated in Fig.~\ref{fig:1d_convergence}, which plots the dispersion $ \omega (0)$ in the thermodynamic limit at momentum $k=0$ as a function of maximum cluster size $N_{\mathrm{max}}$. Both NLCE+VQE and NLCE+ED converge algebraically toward the exact solution, with VQE-ED differences negligible compared to NLCE truncation errors. The excellent agreement between NLCE+VQE and NLCE+ED confirms that $\lceil N/2 \rceil$ HVA layers provide sufficient expressivity for this integrable model. This establishes a proof of principle for the NLCE+VQE approach on an exactly solvable benchmark before proceeding to more challenging cases.

\subsubsection{Square lattice}\label{sec:2D}

The two-dimensional TFIM on a square lattice provides a more stringent test of our framework. In contrast to the one-dimensional case, the 2D TFIM is nonintegrable and lacks an exact analytical solution. 
Nevertheless, extensive quantum Monte Carlo and series-expansion studies have accurately located the quantum critical point at $J_c = 0.328$~\cite{He1990, Hesselmann2016}. 
The transition separates a ferromagnetically ordered phase at large $J$ from a quantum paramagnet polarized along the transverse field at small $J$.

Near the critical point, the system exhibits universal scaling behavior belonging to the $(2+1)$-dimensional Ising universality class. At the quantum critical point, the excitation gap $\Delta(\vec{k})$ closes at momentum $(k_x, k_y) = (0,0)$, resulting in critical fluctuations and a linear dispersion $\omega(\vec{k}) \propto |\vec{k}|$. The vanishing gap at criticality presents a challenging test for NLCE convergence, as larger cluster sizes are required to capture the diverging correlation length. 

Using the NLCE+VQE framework, we compute the 1QP dispersion for the square lattice in the thermodynamic limit at the quantum critical point $J_c = 0.328$, as shown in Fig.~\ref{fig:2d_critical_dispersion}. The figure presents a comparison between dispersions obtained from NLCE+VQE and NLCE+ED for different NLCE orders $N_{\mathrm{max}}$. For these calculations, we employ the rectangular graph expansion defined in Eq.~\ref{eqn:rect_exp} and use $\lceil N/2 \rceil$ layers of the HVA. The energy variance cost function given in Eq.~\ref{eq:C_var} is minimized to obtain the lowest 1QP excitation energy for each cluster. The resulting NLCE+VQE data are compared with ED-based NLCE results and with a high-order series expansion up to 12th order~\cite{He1990}, which we obtained using perturbative continuous unitary transformations~\cite{knetter2000perturbation}, serving as our benchmark. 

Overall, the NLCE+VQE dispersions show very good agreement with both NLCE+ED and the series-expansion benchmark, demonstrating that the VQE-based evaluation of cluster properties can accurately capture the low-energy excitations of the model. The inset highlights the behavior near $(k_x,k_y)=(0,0)$, which corresponds to the point of gap closing at the critical point. In this region, the NLCE convergence becomes increasingly demanding because the excitation gap approaches zero, and larger cluster sizes are required to obtain quantitatively reliable results. Nevertheless, the systematic improvement with increasing $N_{\mathrm{max}}$ indicates that the NLCE+VQE approach is capable of reproducing the correct critical behavior (just as NLCE+ED) when sufficient cluster sizes are included. At $N_{\mathrm{max}}=14$, both NLCE+VQE and NLCE+ED agree to within $\Delta\omega \lesssim 10^{-2}$ at the critical momentum, demonstrating that $\lceil N/2 \rceil$ HVA layers remain sufficient even at criticality for the 2D TFIM. Let us note that scaling of the results in the cluster size is needed to obtain the gapless behavior of the quantum critical point, which is expected due to the diverging correlation length.

\begin{figure}
    \centering
    \includegraphics[width=1\linewidth]{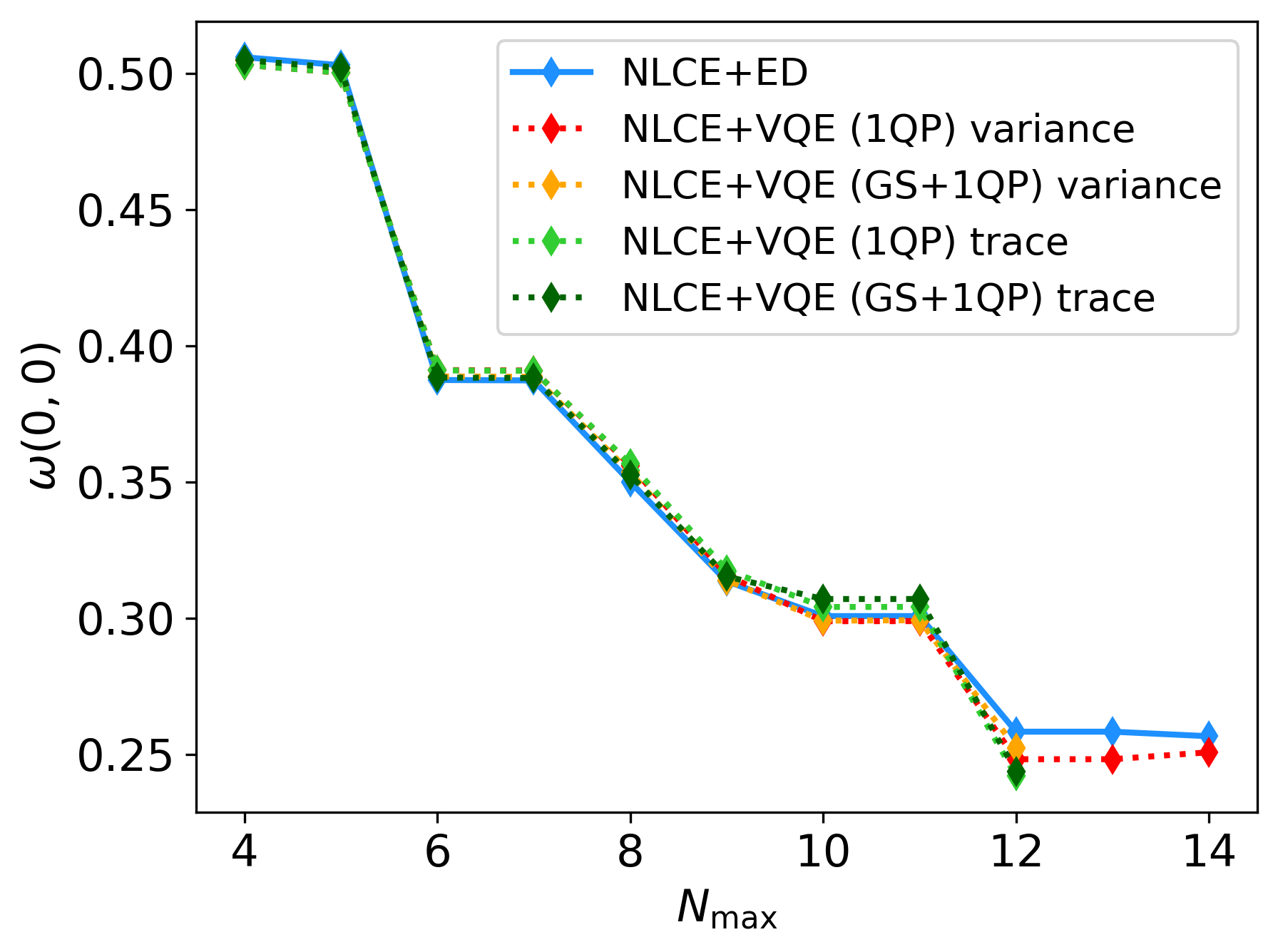}
    \caption{Cost function comparison for the 2D TFIM at the critical point $J_c=0.328$ \cite{He1990, Hesselmann2016}. The plot shows $\omega(0,0)$ for four cost functions: (1QP) variance, (GS+1QP) variance, (1QP) trace, and (GS+1QP) trace. All functions converge identically, with NLCE+ED (blue) shown for reference. The x axis shows the highest cluster $N_{\mathrm{max}}$ used in the NLCE calculation.}
    \label{fig:cfs_comparison_2d}
\end{figure}

Figure~\ref{fig:cfs_comparison_2d} quantifies this convergence behavior, showing that the dispersion $\omega(0,0)$ decreases systematically with increasing $N_{\mathrm{max}}$ for all cost functions tested, confirming both the NLCE convergence and the equivalence of cost function choices at the critical point.

\textit{Cost function comparison.} For the 2D TFIM, we tested four cost functions for the 1QP sector: variance based ($C_{\mathrm{var}}^{\mathrm{1QP}}$, $C_{\mathrm{var}}^{\mathrm{GS,1QP}}$) and trace based ($C_{\mathrm{tr}}^{\mathrm{1QP}}$, $C_{\mathrm{tr}}^{\mathrm{GS,1QP}}$). Figure~\ref{fig:cfs_comparison_2d} shows that all cost functions yield essentially identical NLCE convergence, with variations $\lesssim 10^{-3}$ in the thermodynamic-limit dispersion at $(k_x,k_y)=(0,0)$. Importantly, these cost-function-induced variations are significantly smaller than the NLCE truncation errors (differences between successive $N_{\mathrm{max}}$ values, typically $\sim 5 \times 10^{-2}$) and comparable to or smaller than the VQE-ED differences. However, for the $5\times2$ cluster, variance-based cost functions with ground-state initialization encountered convergence difficulties, while trace-based approaches remained robust. Near-zero initialization resolved this issue. The underlying causes of this initialization sensitivity are analyzed in detail for TFIM+LF (Sec.~\ref{sec:TFIM+LF}).

For the TFIM+LF case (Sec.~\ref{sec:TFIM+LF}), where both parity symmetry breaking and non-integrability make optimization more challenging, we perform a more detailed analysis of initialization sensitivity and its impact on convergence.

\subsection{Transverse-field Ising chain in longitudinal field}{\label{sec:TFIM+LF}}

%\subsubsection{PCAT necessity for TFIM+LF}

\begin{figure}
    \centering
    \includegraphics[width=1\linewidth]{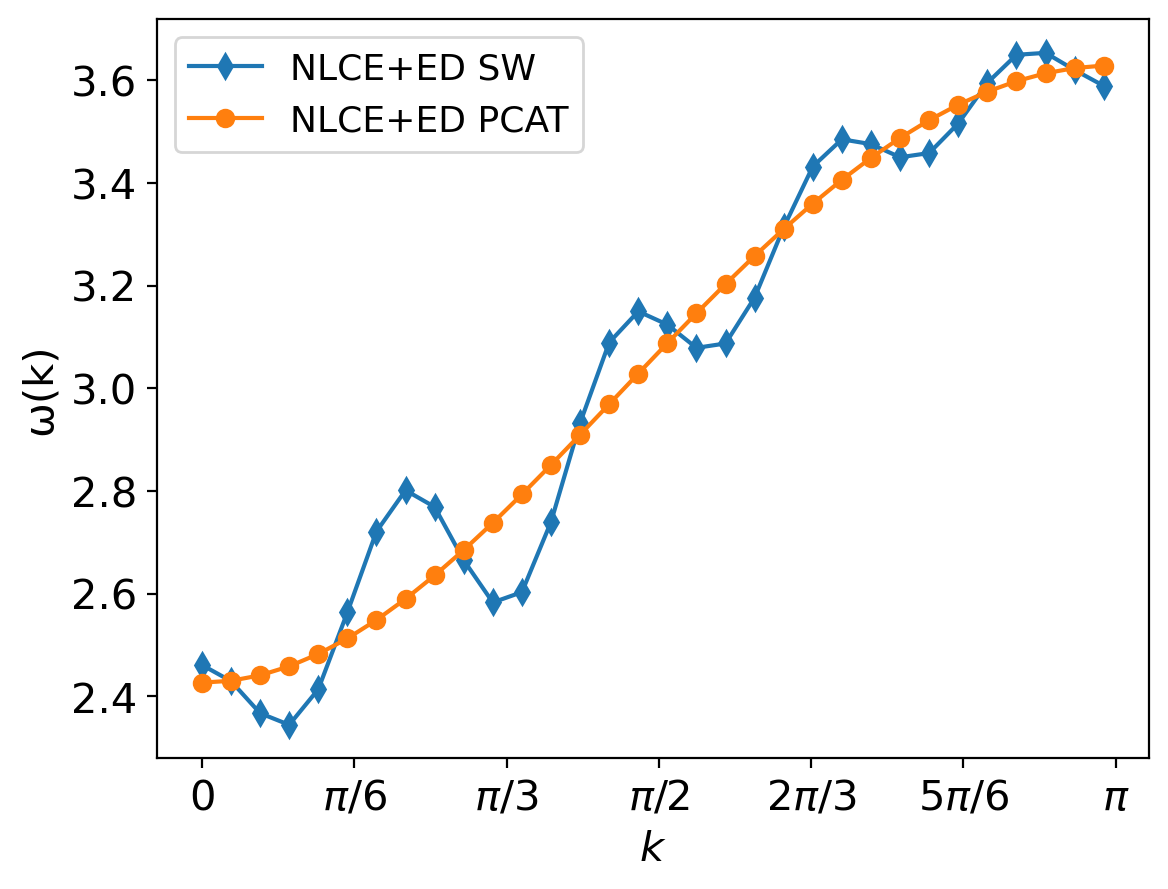}
    \caption{Dispersion $\omega(k)$ for the TFIM+LF obtained with NLCE+ED, where the effective Hamiltonian is constructed using the 2b transformation, compared against the result obtained using PCAT. Results are shown for $N_{\mathrm{max}} = 10$ spins, $J = 0.5$, and $h_l = 0.5$. Blue line: Effective Hamiltonian from 2b transformation alone shows unphysical oscillations due to non-cluster-additive (nonlocal) hopping. Orange line: PCAT-corrected result yields smooth dispersion restoring cluster additivity. }
    \label{fig:sw_pcat_ed}
\end{figure}

The addition of a longitudinal magnetic field $h_l$ to the TFIM fundamentally alters the model's physics. The Hamiltonian becomes
\begin{equation}\label{ham_tfim_lf}
    H = - h \sum_\nu Z_\nu - J \sum_{\langle\nu^\prime,\nu\rangle} X_{\nu^\prime} X_{\nu^{\phantom\prime}} - h_l \sum_\nu X_\nu,
\end{equation}
where the longitudinal field $h_l$ explicitly breaks the $\mathbb{Z}_2$ parity symmetry. This symmetry breaking has important consequences: It couples the ground state to excited states, removes the free-fermion structure that made the pure TFIM chain exactly solvable, and renders the model nonintegrable. We investigate the disordered phase at $J = 0.5$, $h_l = 0.5$ (with $h = 1$).

PCAT becomes crucial for this model, as the 2b transformation alone becomes insufficient due to its allowance of hopping processes between disconnected clusters, as illustrated in Fig.~\ref{fig:hopping}. The longitudinal field induces large overlaps between the unperturbed ground state and the 1QP eigenstates, necessitating careful treatment to not violate cluster additivity. This is explained in detail in Sec.~\ref{sec:PCAT}. The aforementioned issue is further clarified in Fig.~\ref{fig:sw_pcat_ed}, which shows the 1QP dispersion for $N_{\mathrm{max}}=10$ at $J=0.5$ and $h_l=0.5$, obtained using NLCE+ED. The blue solid line in this figure corresponds to the effective Hamiltonian $H_{\mathrm{eff}}$ derived from the 2b transformation, while the orange solid line represents the PCAT result. The PCAT visibly corrects the 2b data, yielding a much smoother dispersion. Under the 2b transformation, a dressed quasiparticle can nonphysically hop between clusters that do not share a bond, violating cluster additivity. PCAT systematically removes these spurious couplings and restores the required additivity condition for NLCE. 
Such nonphysical oscillations become increasingly pronounced with larger cluster sizes and are already significant at $N=9$. Critically, the 2b transformation without PCAT correction fails to produce convergent NLCE results in this regime: The NLCE sum diverges as cluster size increases, making thermodynamic-limit extrapolation impossible. Only after PCAT correction does the NLCE converge, demonstrating that enforcing cluster additivity is essential for obtaining meaningful thermodynamic-limit results.

\begin{figure}
    \centering
    \includegraphics[width=1\linewidth]{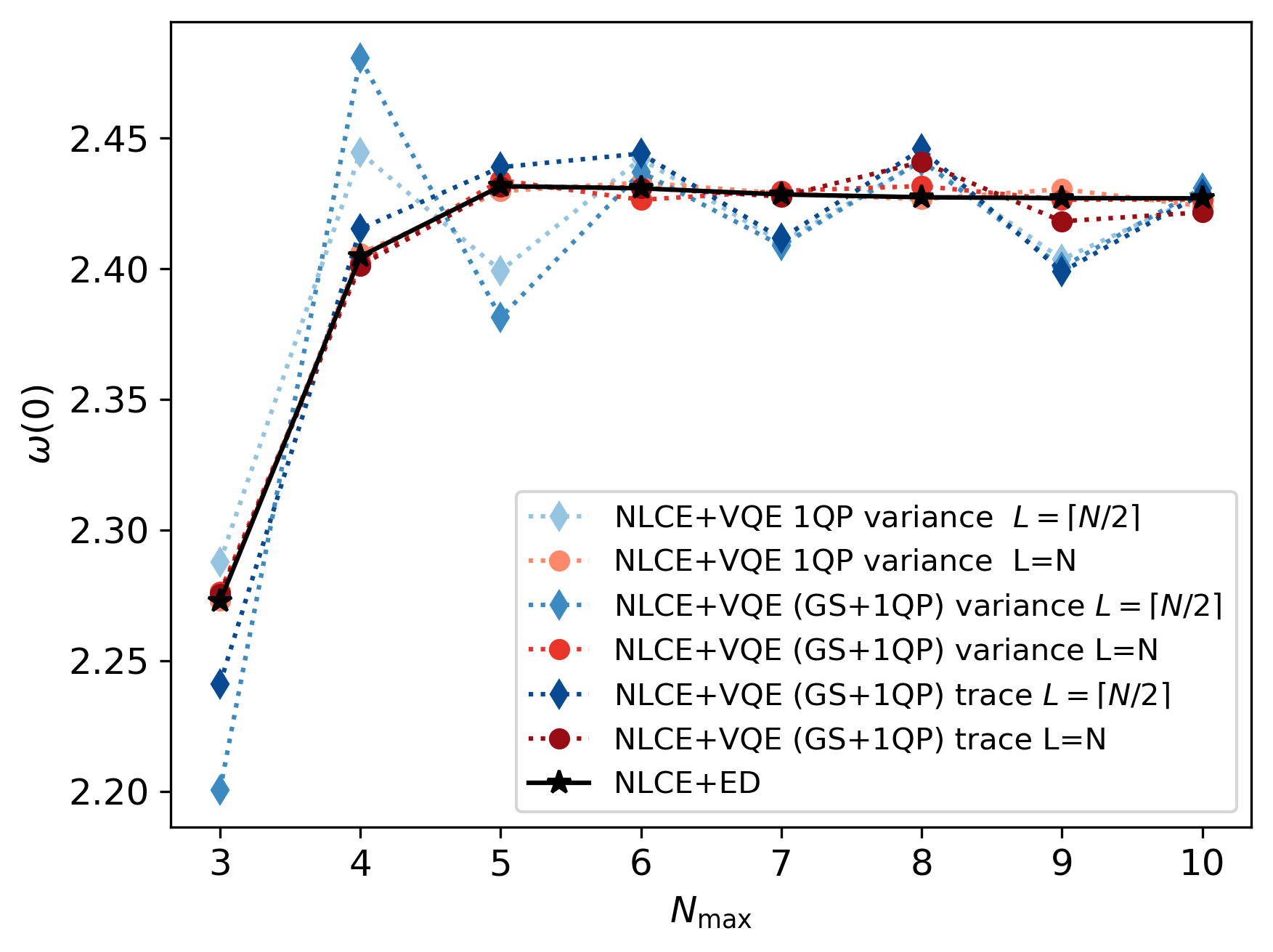}
    \caption{Dispersion at momentum $k = 0$, denoted as $\omega(0)$, in the thermodynamic limit for the TFIM+LF at $J = 0.5$ and $h_l = 0.5$ as a function of the maximum cluster size $N_{\mathrm{max}}$. NLCE+VQE results are shown for all three cost functions using HVA with $\lceil N/2 \rceil$ layers (blue lines with diamonds) and $N$ layers (red lines with circles). The black solid line with stars represents NLCE+ED results.}
    \label{fig:diff_tfim_lf}
\end{figure}

\begin{figure}
    \centering
    \includegraphics[width=1\linewidth]{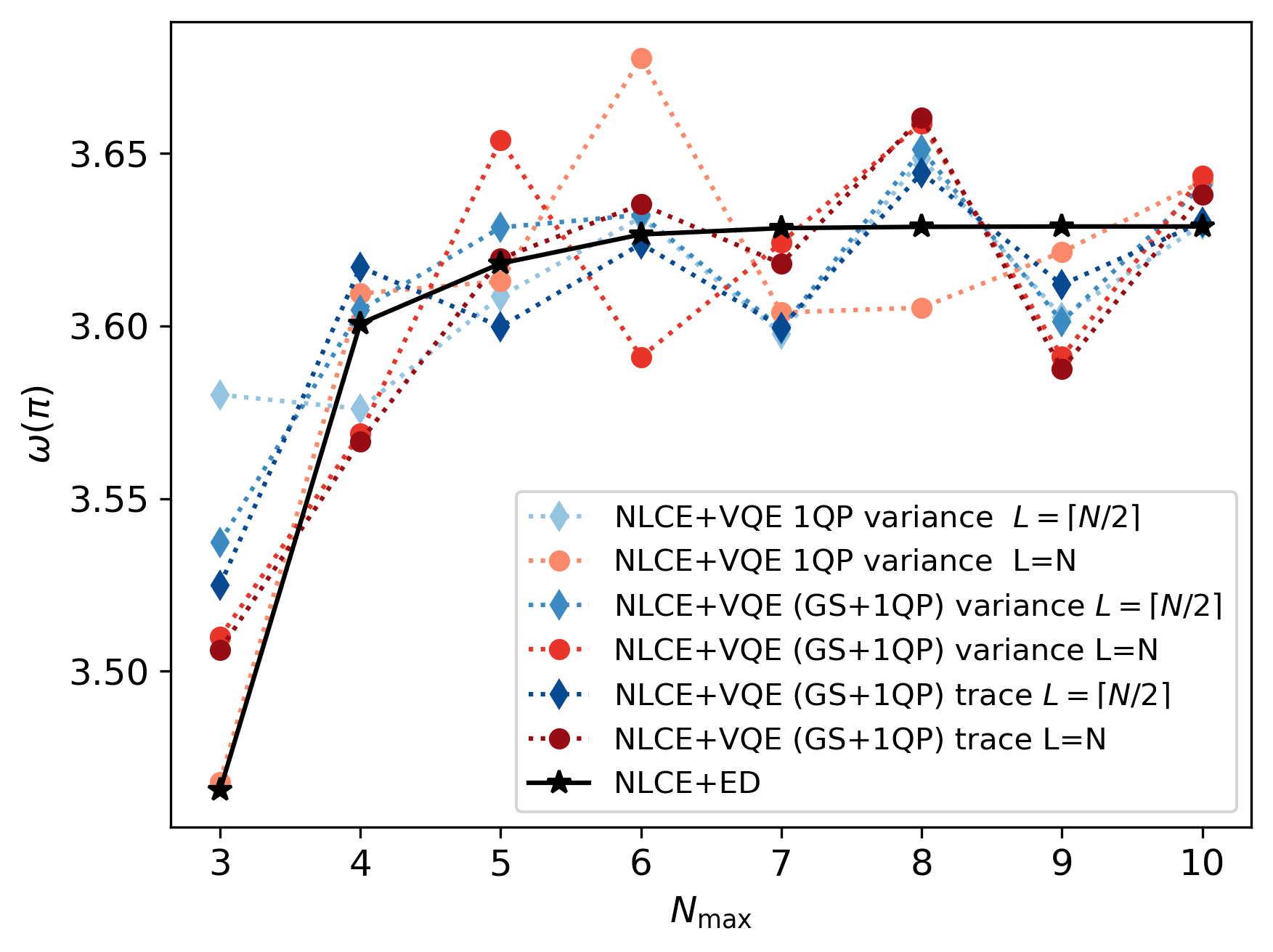}
    \caption{Dispersion at momentum $k = \pi$, denoted as $\omega(\pi)$, in the thermodynamic limit for the TFIM+LF at $J = 0.5$ and $h_l = 0.5$ as a function of the maximum cluster size $N_{\mathrm{max}}$. NLCE+VQE results are shown for all three cost functions using HVA with $\lceil N/2 \rceil$ layers (blue lines with diamonds) and $N$ layers (red lines with circles). The black solid line with stars represents NLCE+ED results.} 
    \label{fig:diff_tfim_lf_pi}
\end{figure}
\subsubsection{Convergence results and cost function analysis}

Since we consider the TFIM+LF on a one-dimensional chain, the NLCE again reduces to the expression in Eq.~\eqref{eq:1d_nlce}. We now perform a detailed comparison of cost functions and initialization strategies for the 1D TFIM+LF. The combination of broken parity symmetry and non-integrability makes this a particularly challenging optimization problem. We investigate convergence speed, robustness to local minima, and layer requirements for different cost functions.

Figures~\ref{fig:diff_tfim_lf} and~\ref{fig:diff_tfim_lf_pi} show NLCE convergence comparing $\lceil N/2 \rceil$ and $N$ layers at both $k=0$ and $k=\pi$ for all three cost functions: variance of 1QP states alone ($C_{\mathrm{var}}^{\mathrm{1QP}}$), variance of combined ground and 1QP states ($C_{\mathrm{var}}^{\mathrm{GS,1QP}}$), and trace of the combined subspace ($C_{\mathrm{tr}}^{\mathrm{GS,1QP}}$). Regarding layer requirements, both $\lceil N/2 \rceil$ and $N$ layers achieve NLCE convergence with increasing cluster size. At $k=0$ (Fig.~\ref{fig:diff_tfim_lf}), $N$ layers provide improved accuracy for intermediate cluster sizes ($N_{\mathrm{max}} \approx 7$--9), though both layer choices converge to comparable accuracy by $N_{\mathrm{max}}=10$. The transient advantage of $N$ layers contrasts with the pure TFIM (Sec.~\ref{sec:2D}) where $\lceil N/2 \rceil$ layers proved sufficient throughout. At $k=\pi$ (Fig.~\ref{fig:diff_tfim_lf_pi}), both layer choices show comparable accuracy throughout, with results converging to NLCE+ED at the largest cluster sizes.

The effective Hamiltonian construction depends on accurately identifying the low-energy eigenspace, which all cost functions accomplish when optimization converges. However, for the most challenging case ($N=10$ spins, $N$ layers), we observed that the initialization strategy using ground-state parameters led to convergence failures for variance-based cost functions, while the trace-based cost function remained robust. Similar behavior was also observed for the $5\times2$ cluster in the 2D TFIM (Sec.~\ref{sec:2D}). For context, successful convergence typically achieves variance values $C_{\mathrm{var}} \sim 0.01$, corresponding to subspace infidelities $1 - \mathcal{F} \sim 10^{-4}$ relative to the target eigenspace, while optimizations trapped in local minima exhibited $C_{\mathrm{var}} \sim 0.5$, indicating substantial contamination from higher excitation sectors. Specifically, $C_{\mathrm{var}}^{\mathrm{1QP}}$ and $C_{\mathrm{var}}^{\mathrm{GS,1QP}}$ became trapped in local minima with cost function values more than 10 times higher than successful optimizations. Initializing all variational parameters near zero resolved this issue for all cost functions, though requiring more optimization iterations due to the less favorable starting point.
\begin{figure}
    \centering
    \includegraphics[width=1\linewidth]{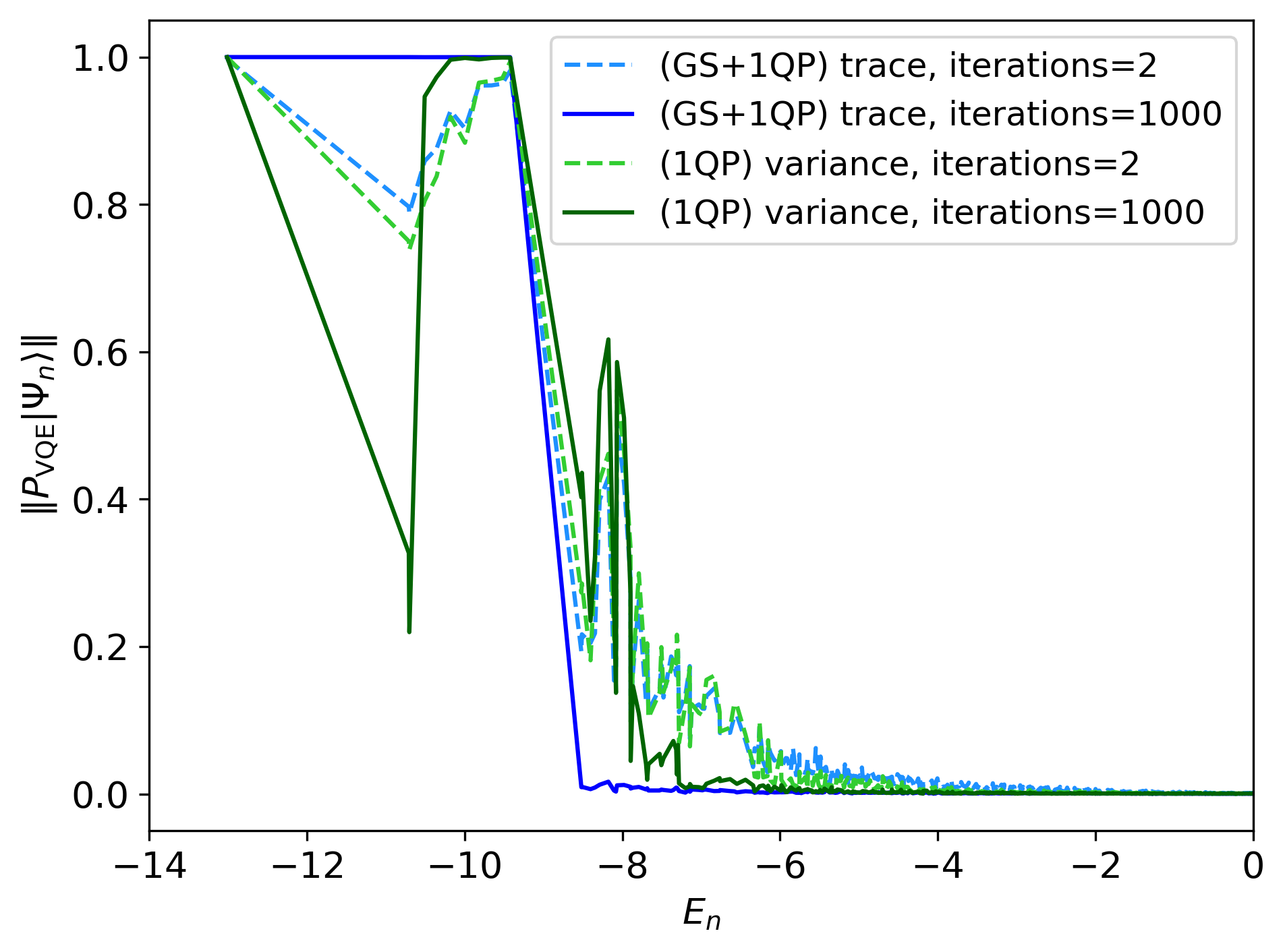}
    \caption{Energy decomposition during VQE optimization for TFIM+LF for $N_{\mathrm{max}}=10$ spins, $J=0.5$, and $h_l=0.5$ using $N$ layers of HVA with conjugate gradient optimization. For selected optimization iterations, we plot the norm $\|P_{\mathrm{VQE}}|\Psi_n\rangle\|$ of the projection of the optimized VQE subspace onto exact energy eigenstates $|\Psi_n\rangle$ as a function of eigenstate energy $E_n$, where $P_{\mathrm{VQE}}$ is the projector onto the optimized subspace. Dashed lines: after 2 conjugate gradient iterations (initial state); solid lines: after 1000 iterations (attempted convergence). Blue lines: Trace cost function $C_{\mathrm{tr}}^{\mathrm{GS,1QP}}$ shows successful convergence with norm values concentrating in ground-state  and 1QP sectors, with negligible projection onto higher excitations. Green lines: variance cost functions $C_{\mathrm{var}}^{\mathrm{1QP}}$ and $C_{\mathrm{var}}^{\mathrm{GS,1QP}}$ show failed convergence with norm values $\|P_{\mathrm{VQE}}|\Psi_n\rangle\| \approx 0.5$ to $0.6$ persisting in 2QP and 3QP sectors (higher-energy region around $E \approx -8$ to $-6$), barely evolving from 2 to 1000 iterations. This indicates the optimizer is trapped in a local minimum.}
    \label{fig:energy_decomposition_tfim_lf}
\end{figure}
The initialization sensitivity reveals important structure in the optimization landscape. Figure~\ref{fig:energy_decomposition_tfim_lf} provides insight for $N=10$ by tracking where the VQE state resides during optimization. We decompose the optimized VQE subspace at each iteration into exact eigenstates (obtained via ED) and plot the norm $\|P_{\mathrm{VQE}}|\Psi_n\rangle\|$ versus eigenstate energy $E_n$, where $P_{\mathrm{VQE}}$ is the projector onto the optimized subspace.

The trace cost function (successful case) shows clean behavior: The norm values concentrate progressively in the ground-state  and 1QP sectors, with negligible projection ($\|P_{\mathrm{VQE}}|\Psi_n\rangle\| < 0.1$, corresponding to $<1\%$ probability weight) onto higher-energy excitations by convergence. The variance cost functions (failed cases) tell a different story: Norm values of $\|P_{\mathrm{VQE}}|\Psi_n\rangle\| \approx 0.5$ to $0.6$ (corresponding to approximately $25\%$ to $35\%$ probability weight) persist in two-quasiparticle (2QP) and three-quasiparticle (3QP) sectors throughout optimization, barely changing from initial to final iterations. The optimizer cannot decouple these higher sectors despite targeting only ground and 1QP states.

This reveals the underlying problem: The two cost functions have different optimization landscapes. Starting from ground-state parameters places the variance optimization in a local minimum where the VQE states are not proper eigenstates but remain superpositions across ground, 1QP, 2QP, and 3QP sectors. While variance vanishes for any eigenstate regardless of which sector it belongs to, here the optimizer fails to find eigenstates altogether; the states remain mixed rather than collapsing to eigenstates of the full Hamiltonian. The trace cost function avoids this trap, likely because direct energy minimization provides stronger gradients away from such mixed configurations than variance minimization.

\begin{figure}
    \centering
    \includegraphics[width=1\linewidth]{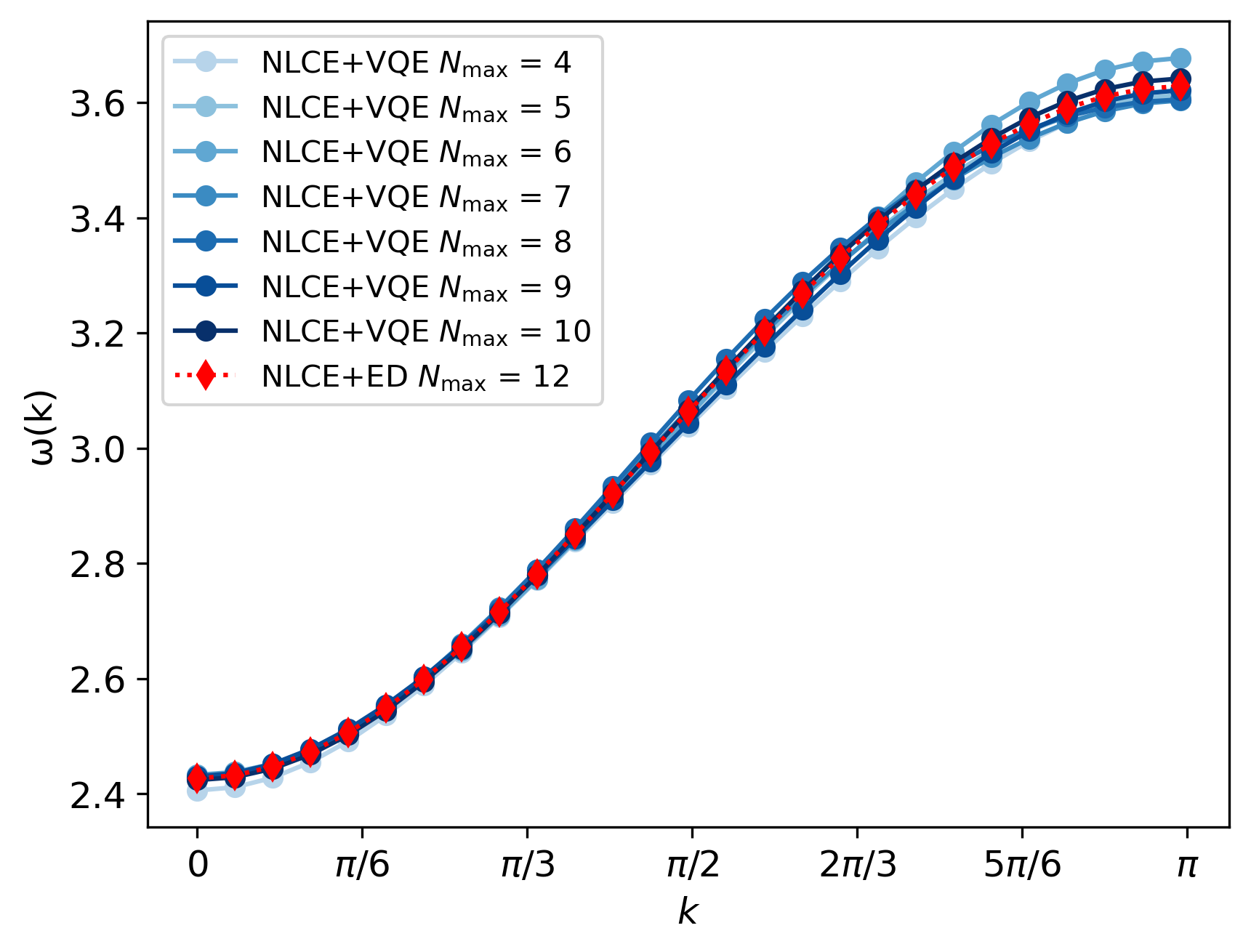}
    \caption{Dispersion $\omega(k)$ in the thermodynamic limit for TFIM+LF at $J =0.5$ and $h_l=0.5$. Blue lines show NLCE+VQE results for different maximum cluster sizes up to $N_{\mathrm{max}}=10$, while the red dotted line shows NLCE+ED for $N_{\mathrm{max}}=12$. VQE uses $N$ layers of HVA with variance $C_{\mathrm{var}}^{\mathrm{GS,1QP}}$ cost function.}
    \label{fig:dispersion_tfim_lf}
\end{figure}

In Fig.~\ref{fig:dispersion_tfim_lf}, we show the 1QP dispersion $\omega(k)$ in the thermodynamic limit obtained from NLCE+VQE (using $N$ layers and variance $C_{\mathrm{var}}^{\mathrm{GS,1QP}}$ with near-zero initialization) compared to NLCE+ED. NLCE+VQE converges systematically toward NLCE+ED with increasing maximum cluster size $N_{\mathrm{max}}$. Convergence is slower than for both the pure 1D TFIM and the 2D TFIM cases. This reflects the increased optimization difficulty: The longitudinal field breaks integrability and parity symmetry, coupling odd and even quasiparticle sectors and creating a more complex optimization landscape than in the pure TFIM cases. The interplay between optimization difficulty and physical correlation structure requires further investigation: The longitudinal field may increase entanglement within clusters, or the slower NLCE convergence may stem primarily from the harder optimization problem rather than from a genuinely longer correlation length. Determining the relative contributions of these effects remains an open question.

\section{Conclusions}\label{sec:conclusion}

In this work, we have developed a hybrid quantum-classical method for computing quasiparticle excitation energies in the thermodynamic limit, introducing a quantum algorithm capable of constructing cluster-additive effective Hamiltonians for excited states. The key challenge is ensuring cluster additivity for degenerate subspaces, a property not automatically satisfied by variational optimization. This is resolved by integrating the PCAT \cite{hormann_projective_2023} with VQE. This integration is non-trivial: VQE provides the essential information (energy expectations and state overlaps) needed to construct PCAT without exponentially costly state tomography, enabling NLCE convergence for excited states with quantum algorithms as cluster solvers.

Our benchmarks on the TFIM demonstrate that VQE with the HVA achieves convergence using $\lceil N/2 \rceil$ layers for the pure TFIM in both one and two dimensions. For the TFIM with longitudinal field, where parity symmetry breaking makes PCAT essential, the situation becomes more nuanced: While $\lceil N/2 \rceil$ layers show convergence with increasing cluster size, using $N$ layers provides improved accuracy. 

For the variance-based cost function, we observe that near-zero initialization outperforms the ground-state-initialized strategy, while the trace cost function remains robust to ground-state initialization. This reveals that the two cost functions have qualitatively different optimization landscapes: The variance landscape contains local minima near ground-state parameters that trap the optimizer, while the trace landscape does not. Cluster sizes beyond $N \sim 10$ become challenging for this combination of variance cost function with suboptimal initialization.

The PCAT framework provides several advantages beyond ensuring cluster additivity. By constructing the transformation from measured energy expectations and state overlaps, it requires $\mathcal{O}(N^2)$ measurements per cluster, polynomial overhead that scales favorably compared to exponential state space growth. 

Beyond the specific application to quasiparticle excitations, this work demonstrates that the PCAT framework extends naturally to thermodynamic-limit calculations through appropriate enforcement of cluster additivity. Importantly, while we have focused on VQE implementation, the PCAT construction is not intrinsically tied to variational optimization. The framework relies on measuring energy expectations and state overlaps, as detailed in Sec.~\ref{sec:measurements}, making it immediately ready for application when fault-tolerant quantum protocols achieve quantum advantage. The same measurement protocol applies equally to quantum phase estimation, adiabatic state preparation, and quantum annealing. This algorithmic flexibility positions the approach to scale with advancing quantum hardware and ensures it remains relevant even if variational methods face fundamental limitations such as barren plateaus in certain regimes. The measurement protocol thus provides a practical pathway for implementation across different quantum algorithmic paradigms.

Implementation of NLCE+VQE with PCAT on real quantum hardware is currently underway to assess performance under realistic noise and finite measurement sampling. We will investigate how gate errors and shot noise propagate through the PCAT construction and subsequent NLCE summations. To mitigate error accumulation, the hybrid classical-quantum nature suggests a strategy where the largest clusters, those contributing most to NLCE convergence but being classically intractable, are treated with VQE, while smaller clusters use exact diagonalization to minimize accumulated errors.

Several extensions present themselves naturally. Computing two quasiparticle sectors would enable investigation of bound states and scattering continua, though the quadratically growing subspace dimensions make this significantly more challenging. Dynamical structure factors, the natural extension of static structure factors to finite-frequency response, are directly accessible through our framework by evaluating local observables between the ground state and quasiparticle excitation states constructed via NLCE. These spectral functions are directly measurable in inelastic neutron or light scattering experiments and provide detailed information about the elementary excitation spectrum. The framework's basis-independent formulation makes it applicable to various quantum lattice models, though we note that gapped systems where quasiparticle excitations are well defined represent the most natural domain of applicability given the current formalism.

For conventional NLCE with classical algorithms, the logarithm $S = \ln T$ of the transformation possesses cluster-additive structure when properly constructed \cite{hormann_projective_2023}. Whether analogous principles can be established for constructing approximate unitaries for larger clusters from solutions on smaller constituent subgraphs in variational circuits, potentially extending beyond HVA to other ansätze, remains an open question that could improve robustness to optimization errors and help circumvent local minima.

The challenges encountered already with TFIM with longitudinal field (where optimization becomes difficult) underscore the importance of systematic investigation across diverse lattice models. Future applications to frustrated quantum magnets represent a long-term vision where quantum approaches might prove particularly valuable: Frustrated systems suffer from severe sign problems that make quantum Monte Carlo exponentially harder with system size, rendering many frustrated systems practically inaccessible. While NLCE converges more slowly for frustrated systems and requires larger clusters, quantum algorithms can potentially access these larger cluster sizes where classical exact diagonalization fails. The combination of NLCE with quantum algorithms could thus provide access to parameter regimes where classical methods cannot reach, offering a domain where quantum advantages emerge naturally. Establishing the approach's general applicability and practical utility across different model classes remains essential.

\section*{Acknowledgments}
 Sumeet and M. H. acknowledge Lucas Marti for discussions on quantum hardware application. We acknowledge the support by the Munich Quantum Valley, which is supported by the Bavarian state government with funds from the Hightech Agenda Bayern Plus.

\textbf{Data Availability:} The data that support the findings of this article are openly available in Zenodo~\cite{data}.

\nocite{*}

% Produces the bibliography via BibTeX.

\end{document}